\documentclass[lettersize,journal]{IEEEtran}

\usepackage{amsmath,amsfonts}
\usepackage{amssymb}
\usepackage{algorithm}
\usepackage{algpseudocode}
\usepackage{array}
\usepackage{booktabs}
\usepackage{tabularx}
\usepackage{makecell}
\usepackage{textcomp}
\usepackage{stfloats}
\usepackage{url}
\usepackage{verbatim}
\usepackage{graphicx}
\usepackage{cite}
\usepackage{tikz}
\usetikzlibrary{arrows.meta,positioning,fit,calc,shapes.geometric}

\begin{document}

\title{SFC-Aware Online Aggregated Data-Link Orchestration for SDN/NFV-Enabled SAGINs}

\author{Ziyang~Guo and Bing~Du,~\IEEEmembership{Member,~IEEE}%
\thanks{Ziyang Guo and Bing Du are with the Department of Computer and Communication Engineering, University of Science and Technology Beijing, Beijing, China (e-mail: m202420965@xs.ustb.edu.cn; dubing@ustb.edu.cn).}%
\thanks{This work was supported by the National Natural Science Foundation of China under Grant 62471031.}%
\thanks{Corresponding author: Bing Du (e-mail: dubing@ustb.edu.cn).}}

\markboth{IEEE Transactions on Vehicular Technology}%
{Guo and Du: SFC-Aware Online Aggregated Data-Link Orchestration for SDN/NFV-Enabled SAGINs}

\maketitle
\begin{abstract}
Civil aviation space-air-ground integrated networks (SAGINs) are expected to support heterogeneous cockpit and cabin services over dynamic air-to-air (A2A), air-to-ground (A2G), and air-to-satellite (A2S) data links. Existing static data-link binding schemes cannot fully exploit such heterogeneous communication opportunities, while repeatedly solving mixed-integer orchestration models is difficult for fine-grained online control. This paper studies service function chain (SFC)-aware online access-side aggregated data-link orchestration for civil aviation SAGINs enabled by software-defined networking and network function virtualization (SDN/NFV). The key idea is to jointly orchestrate spatial bearer resources provided by heterogeneous A2A/A2G/A2S data links and temporal elasticity enabled by temporal elastic mapping and parking (TEMP). A rolling-slot optimization model is developed with four request-level actions: NOW, TEMP, REJECT, and DROP. The model follows a lexicographic objective that prioritizes service success under quality-of-service (QoS) deadlines, then normalized access-orchestration delay, and finally residual TEMP-related risk. To avoid exhaustive search over the full binary action space, we propose Model-Induced Risk and Scarcity-Aware Refinement (MRSAR), a model-induced online approximation algorithm. MRSAR derives deferred-realization risk from the success-prioritized objective and extracts resource scarcity signals from data-link and TEMP buffer constraints. These model-induced signals are used for valuation-based feasible solution construction and bounded neighborhood refinement, thereby deciding whether each request should be served immediately, preserved for future realization, rejected, or dropped. Simulation results show that MRSAR remains close to the Gurobi-mixed-integer linear programming (MILP) reference in service success, outperforms arrival-order and delay-myopic greedy baselines, controls REJECT and DROP failures, and achieves a favorable quality-complexity tradeoff for rolling online orchestration.
\end{abstract}

\section{Introduction}

The rapid growth of civil aviation services is reshaping the role of aeronautical communication networks. Modern aircraft are expected to support cockpit safety services, airline operation services, bandwidth-intensive cabin services, and delay-sensitive interactive applications with heterogeneous traffic volumes, delay budgets, and reliability requirements. Space-air-ground integrated networks (SAGINs) provide a promising architecture for extending communication coverage and improving service continuity by integrating satellite, aerial, and terrestrial segments \cite{liu2018sagin,cheng2022service_sagin,zhou2023aerospace,azari2022ntn}. In civil aviation scenarios, such integration naturally involves heterogeneous air-to-air (A2A), air-to-ground (A2G), and air-to-satellite (A2S) data links, whose capacity, propagation delay, availability, and service suitability may vary rapidly with aircraft mobility and network load \cite{bilen2022aeronautical}. Request-level online orchestration over these links is therefore a nontrivial problem.

In civil aviation SAGINs, aircraft should not be regarded only as traffic sources. They can also serve as a programmable intermediate layer that connects heterogeneous spatial communication opportunities. From this perspective, A2A, A2G, and A2S data links can be abstracted as aggregated data-link bearer resources and jointly managed by a network control plane. Software-defined networking (SDN) and network function virtualization (NFV) provide the architectural basis for such abstraction, since SDN separates control logic from forwarding behavior and NFV decouples network functions from dedicated hardware \cite{kreutz2015sdn,mijumbi2016nfv}. Network slicing and softwarization further support flexible resource sharing in virtualized networks \cite{afolabi2018network_slicing}. This abstraction is also consistent with recent SAGIN resource-management studies that emphasize unified control over heterogeneous space, air, and ground resources \cite{liang2024resource_sagin,zhang2022multi_domain_sagin,zhang2024ai_sagin}. With these capabilities, an aircraft service request can be dynamically assigned to a suitable access anchor and bearer mode before entering a service-function-chain-capable domain.

Service function chain (SFC) steers traffic through an ordered set of service functions and has been widely studied in SDN/NFV-enabled networks \cite{halpern2015sfc_arch,bhamare2016sfc_survey,hantouti2020sfc_5gb}. Recent works have further investigated SFC mapping, embedding, and scheduling in SAGINs and satellite-ground integrated networks \cite{li2022cost_sfc_sagin,zhang2022sfc_sagin,zhang2023fl_sfc_sagin,jia2025sfc_scheduling}. In large-scale and dynamic aeronautical scenarios, however, recomputing the complete service chain for every request in every short decision slot may introduce excessive online overhead. The more urgent decision is often located at the access side: the controller must jointly select a SAGIN access anchor and an
aggregated data-link bearer mode, and determine whether the request
should be served immediately, temporarily preserved, or released under
strict quality-of-service (QoS) and capacity constraints. Here, the access anchor is a selectable SAGIN-side admission entity,
such as an aircraft, satellite node, or ground gateway, while the
bearer mode specifies the associated A2A/A2G/A2S data-link technology. Therefore, this paper focuses on SFC-aware access orchestration rather than full end-to-end SFC remapping. This scope retains service awareness while keeping the online decision process suitable for fine-grained rolling control.

A key difficulty in this problem is that service success depends on both spatial and temporal resources. Spatially, multiple heterogeneous data-link opportunities may be available in a given slot. Temporally, a request that cannot be efficiently served immediately may still be valuable if future link opportunities can realize it within its access-orchestration delay budget. Similar time-varying opportunities have been observed in delay-tolerant, minimum-delay, and multi-objective aeronautical networking studies \cite{du2021dynamic_graph,cui2021minimum_delay_aanet,zhang2022multiobjective_aanet}. Motivated by this observation, this paper introduces temporal elastic mapping and parking (TEMP). TEMP is not treated as a passive waiting queue, but as a schedulable temporal resource supported by onboard storage or storage-type virtualized functions. A request should enter TEMP only when the finite look-ahead window indicates sufficient realization opportunities and when its residence time and buffer occupation remain feasible. The orchestration problem is thus not simply whether to use TEMP, but how to select, preserve, and realize TEMP requests without turning deferred service into delayed failure.

Several existing decision principles are insufficient for this coupled spatial-temporal orchestration problem. Static data-link binding cannot adapt to rapid changes in traffic demand and link availability. Arrival-order scheduling, represented by first-come-first-served service disciplines, is simple to implement \cite{kleinrock1975queueing,ojijo2020slice_admission}, but it does not distinguish requests with different delay budgets, future realization probabilities, or resource impacts. Delay-oriented selection, related to minimum-delay routing and shortest-delay decision principles \cite{gallager1977minimum_delay,cui2021minimum_delay_aanet,zhang2022multiobjective_aanet}, can reduce instantaneous delay, but it is myopic with respect to future TEMP realization and heterogeneous resource scarcity. General-purpose mixed-integer linear programming solvers can provide high-quality references, but repeatedly solving a large binary decision problem in short online slots is computationally expensive \cite{gurobi2026}. Learning-based resource management has also attracted attention in network slicing and aeronautical routing \cite{li2018drl_slicing,liu2021drl_aanet}. Although such methods can reduce post-training decision latency, their black-box nature and indirect constraint handling are less desirable for safety-sensitive aeronautical communication systems, where the controller should be able to explain why a request is served, deferred, rejected, or dropped.

To address these issues, this paper studies SFC-aware online aggregated data-link orchestration for SDN/NFV-enabled civil aviation SAGINs. In each rolling decision slot, the active requests include both newly arrived requests and requests preserved in TEMP from previous slots. For each request, the controller chooses one action from NOW, TEMP, REJECT, and DROP. NOW assigns the request to an immediate feasible access anchor and aggregated bearer mode. TEMP keeps the request for possible future realization under a finite look-ahead window. REJECT handles newly arrived requests that cannot be admitted, while DROP handles carried-over TEMP requests that can no longer be legally preserved. The finite-horizon information used by TEMP is treated as an input to orchestration; in the simulations, recorded future slots are used as look-ahead information to isolate the value of TEMP-aware decision making, while dedicated traffic or link-state prediction is left outside the main focus of this paper.

Based on this action structure, we formulate a success-prioritized rolling optimization model. The model captures immediate data-link capacity, TEMP buffer capacity, TEMP residence limits, future realization probability, and QoS delay constraints. Its objective is organized lexicographically: the first layer maximizes service success, the second layer minimizes normalized access-orchestration delay after protecting success, and the third layer controls residual TEMP-related risk, including DROP, residence urgency, and buffer pressure. This structure avoids collapsing heterogeneous operational priorities into a single weighted-sum objective, and reflects that service success should be protected first, delay should be optimized second, and residual TEMP risk should be reduced only without damaging the preceding objectives.

The resulting mixed-integer model is useful for clarifying the orchestration logic, but it is too expensive for repeated online execution at large request scales. We therefore propose Model-Induced Risk and Scarcity-Aware Refinement (MRSAR), a model-induced online approximation algorithm. MRSAR is not designed as an independent greedy rule; it derives deferred-realization risk from the success-prioritized objective, extracts scarcity-aware signals from current data-link and TEMP buffer constraints, and refines feasible actions within constrained NOW/TEMP/REJECT/DROP neighborhoods. In this way, MRSAR preserves the interpretability of the original model while avoiding exhaustive search over the full binary action space.

The main contributions of this paper are as follows.

\begin{itemize}
\item We formulate an SFC-aware online aggregated data-link orchestration problem for SDN/NFV-enabled civil aviation SAGINs, where A2A, A2G, and A2S data links are treated as heterogeneous aggregated data-link bearer resources for request-level access anchor and bearer-mode selection.

\item We introduce TEMP as a temporal elastic mapping and parking mechanism. TEMP transforms onboard storage from passive buffering into a schedulable temporal resource whose value depends on future realization probability, delay feasibility, residence state, and buffer pressure.

\item We develop a rolling-slot lexicographic optimization model over NOW, TEMP, REJECT, and DROP actions. The model prioritizes QoS success, then normalized access-orchestration delay, and finally residual TEMP-related risk, thereby reflecting service-oriented decision priorities more explicitly than conventional weighted-sum formulations.

\item We propose MRSAR, a model-induced risk and scarcity-aware refinement algorithm. MRSAR uses risk values and resource scarcity signals derived from the optimization model to obtain interpretable online decisions with much lower computational cost than direct MILP solving.

\item We conduct extensive simulations under heterogeneous civil aviation SAGIN settings. The results show that MRSAR remains close to the Gurobi-MILP reference in QoS success, outperforms arrival-order and delay-myopic greedy baselines, controls REJECT and DROP failures, and achieves better runtime scalability for rolling online orchestration.
\end{itemize}

The remainder of this paper is organized as follows. Section~II reviews related work. Section~III presents the system architecture and problem overview. Section~IV formulates the mathematical model. Section~V introduces the MRSAR algorithm and analyzes its complexity. Section~VI reports the simulation results. Section~VII concludes the paper.
\section{Related Work}

\subsection{Resource Orchestration in SAGIN and Aeronautical Networks}

Space-air-ground integrated networks have been widely studied for extending network coverage and supporting heterogeneous services through the integration of satellite, aerial, and terrestrial segments \cite{liu2018sagin,cheng2022service_sagin,zhou2023aerospace,azari2022ntn}. Existing studies have considered resource allocation, network slicing, service coverage, and cross-domain coordination in SAGINs \cite{liang2024resource_sagin,zhang2022multi_domain_sagin,zhang2024ai_sagin}. Aeronautical network studies further show that aircraft networks face highly dynamic topology, heterogeneous data-link characteristics, and stringent service requirements \cite{bilen2022aeronautical,baltaci2021aerial_networks}. These works provide an important basis for resource management in dynamic space-air-ground and aeronautical communication environments.

However, most existing SAGIN orchestration studies focus on general satellite-aerial-terrestrial integration, UAV-assisted networking, routing, or segment-level resource management. The civil aircraft layer is usually treated as a traffic source or a mobile access node, rather than as a programmable intermediate layer that can aggregate heterogeneous A2A, A2G, and A2S data-link opportunities. Delay-aware and multi-objective aeronautical routing studies optimize path selection under dynamic connectivity \cite{cui2021minimum_delay_aanet,zhang2022multiobjective_aanet}, but they do not directly address request-level access anchor selection, bearer-mode selection, and TEMP-based temporal preservation under a unified rolling orchestration model. In contrast, this paper treats A2A, A2G, and A2S links as aggregated data-link bearer resources and studies request-level online orchestration over these resources.

\subsection{SDN/NFV-Enabled SFC Orchestration}

SDN and NFV provide key support for flexible network control and virtualized service deployment \cite{kreutz2015sdn,mijumbi2016nfv}. Based on these techniques, SFC steers traffic through an ordered set of service functions, and existing studies have investigated virtual network function (VNF) placement, chain embedding, and end-to-end service path construction \cite{halpern2015sfc_arch,bhamare2016sfc_survey,hantouti2020sfc_5gb}. Recent works have further extended SFC orchestration to SAGINs and satellite-ground integrated networks, including cost-aware SFC mapping, SFC-based resource allocation, federated-learning-assisted SFC embedding, and dynamic SFC scheduling under time-varying resources \cite{li2022cost_sfc_sagin,zhang2022sfc_sagin,zhang2023fl_sfc_sagin,jia2025sfc_scheduling}.

Nevertheless, directly applying full SFC remapping to fine-grained civil aviation online control is difficult. In a large-scale rolling scenario, recomputing the complete service chain for every active request may introduce excessive decision overhead. Moreover, the immediate bottleneck considered in this paper is not the internal processing order of the service chain, but the access-side decision before a request enters an SFC-capable service domain. Therefore, this paper focuses on SFC-aware access orchestration, where the controller selects the access anchor, aggregated data-link bearer mode, and request-level action while avoiding unnecessary full-chain remapping at every decision slot.

\subsection{Online Solvers, Greedy Policies, and Learning-Based Methods}

Online orchestration problems are commonly addressed by optimization-based methods, low-complexity greedy policies, or learning-based approaches. Mixed-integer linear programming (MILP) can naturally model binary action choices, capacity constraints, and QoS constraints, and commercial solvers such as Gurobi can provide high-quality reference solutions \cite{gurobi2026}. However, repeatedly solving a large mixed-integer model in short rolling slots is computationally expensive, especially when the number of active requests and candidate data-link actions increases. Thus, direct solver-based optimization is more suitable as a performance reference than as an online controller-side mechanism.

Greedy policies are attractive because of their low implementation complexity. Arrival-order service disciplines, such as first-come-first-served scheduling, are widely used in queueing and online service systems \cite{kleinrock1975queueing}. Similar admission-oriented and myopic baselines are also common in dynamic wireless resource optimization and network slicing studies \cite{ojijo2020slice_admission}. In the considered problem, arrival-order processing assigns feasible actions according to request order, while delay-myopic selection greedily chooses the currently shortest feasible service opportunity, following minimum-delay decision principles \cite{gallager1977minimum_delay,cui2021minimum_delay_aanet}. These policies represent service-order simplicity and instantaneous delay minimization, but neither explicitly evaluates TEMP realization opportunities nor prices the scarcity of heterogeneous data-link and buffer resources.

Learning-based resource management has also attracted attention in network slicing and aeronautical networking \cite{li2018drl_slicing,liu2021drl_aanet}. However, such methods usually require sufficient training data and careful reward design. Their decision logic is less transparent than model-based methods, and hard constraints are often handled indirectly through penalties or post-processing. In safety-sensitive aeronautical networking, an online orchestration mechanism should preferably remain interpretable and constraint-aware. Motivated by these observations, this paper develops MRSAR as a model-induced online approximation algorithm. It is designed to be faster than repeated MILP solving, more orchestration-aware than arrival-order and delay-myopic greedy policies, and more interpretable than black-box learning-based methods.

\begin{figure*}[!t]
\centering
\includegraphics[
    width=0.92\textwidth,
    height=0.32\textheight,
    keepaspectratio
]{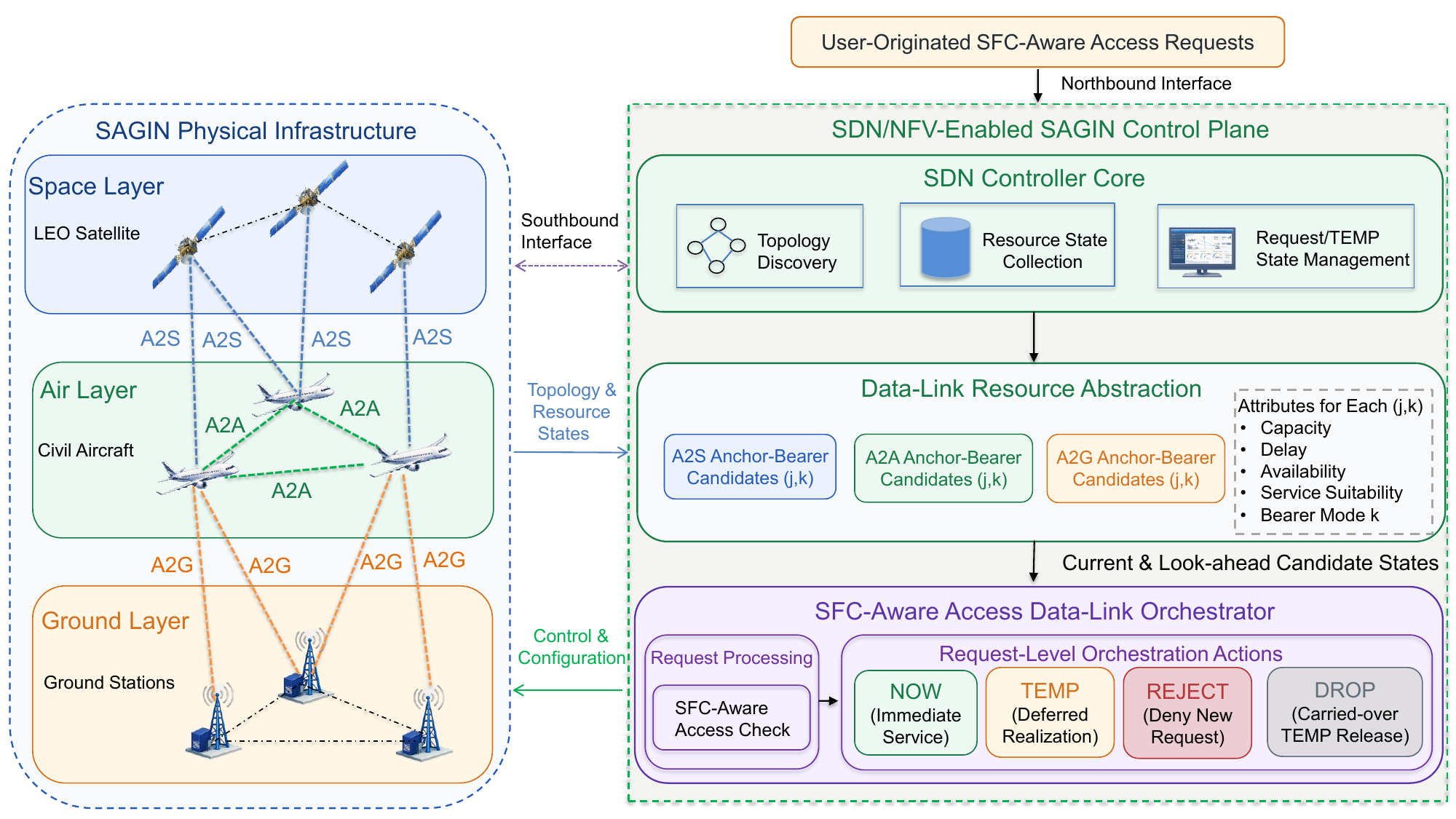}
\caption{Access-side architecture for SFC-aware data-link aggregation in an
SDN/NFV-enabled SAGIN. The SDN controller collects topology and resource
states from the space, air, and ground segments, abstracts A2A/A2G/A2S
links into anchor--bearer candidates, and supports request-level NOW,
TEMP, REJECT, and DROP decisions within the access-orchestration stage.}
\label{fig:system_architecture}
\end{figure*}
\section{System Architecture and Problem Overview}
\label{sec:system_overview}

\subsection{SDN/NFV-Enabled Civil Aviation SAGIN}

We consider an SDN/NFV-enabled civil aviation SAGIN where aircraft, ground gateways, and satellite nodes jointly provide heterogeneous communication opportunities for cockpit and cabin services. As shown in Fig.~\ref{fig:system_architecture}, the aircraft layer is not treated merely as a set of traffic sources. Instead, aircraft are regarded as programmable airborne nodes that can access and relay service requests through A2A, A2G, and A2S data links. The ground segment provides gateways and terrestrial service access, while the satellite segment extends wide-area coverage when terrestrial connectivity is unavailable or congested.

SFC-aware access requests enter the control plane through the northbound interface, while topology, link availability, bearer capacity, delay attributes, and TEMP-related states are collected from the SAGIN infrastructure through the southbound interface. Based on these states, the SDN/NFV controller determines request-level orchestration actions in a rolling-slot manner and translates the decisions into control and configuration commands for selected access anchors, bearer modes, and TEMP storage functions. NFV enables service functions such as storage, protocol conversion, gateway access, traffic filtering, and security processing to be deployed as virtualized components, forming an SFC-capable service domain into which aircraft service requests are admitted through selected access anchors.

This framework differs from conventional static data-link access, where a service class is usually associated with a preferred communication technology or a fixed access mode. Here, heterogeneous A2A, A2G, and A2S data links are abstracted as aggregated data-link bearer resources and jointly orchestrated by the SDN/NFV control plane.

\subsection{Aggregated Data-Link Bearer Resources and Access-Anchor Abstraction}

In the considered civil aviation SAGIN, a request may be carried by different types of data links. A2A links provide airborne cooperation opportunities among aircraft, A2G links connect aircraft to ground stations or gateways, and A2S links provide satellite-based access when ground coverage is limited or satellite connectivity better matches the service requirements. These links differ in transmission rate, propagation delay, resource availability, and supported service types.

Rather than modeling each physical technology as an isolated resource, this paper abstracts heterogeneous data-link opportunities into aggregated data-link bearer resources. Let \(j \in \mathcal{V}\) denote a candidate SFC access anchor,
namely a selectable SAGIN-side admission entity such as an aircraft,
satellite node, or ground gateway. Let \(k \in \mathcal{K}\) denote
an aggregated data-link bearer mode, which represents the selected
A2A/A2G/A2S access category or a concrete bearer technology, including
VDL-Mode-2, HFDL, legacy SATCOM, SBB-Safety, ATG, 5G-ATG, and LEO
broadband. The ordered pair \((j,k)\) represents an anchor--bearer
candidate, meaning that request \(r\) is admitted through access anchor
\(j\) using bearer mode \(k\).

For each anchor--bearer candidate, the controller maintains attributes such as available capacity, transmission rate, delay, availability, and service suitability. These attributes are used to screen current and look-ahead candidate states according to reachability, service compatibility, capacity, and access-delay feasibility. The resulting candidates are later represented by the NOW candidate set and the predicted future candidate sets in the mathematical model. This abstraction avoids statically binding a service class to a single data-link technology and allows heterogeneous communication opportunities to be handled through a unified orchestration view.

The proposed framework is SFC-aware, but it does not recompute a complete service chain for every request at every decision slot. In a large-scale rolling scenario, the sequence of service functions can be maintained by service templates or preconfigured service domains according to the request type. The online bottleneck is the access-side decision, namely how to select an appropriate access anchor and data-link bearer so that the request can enter a suitable SFC-capable service domain under access-delay and capacity constraints. Therefore, an access anchor is not merely a conventional next-hop node, but an entry point toward an SFC-capable service domain. Once a request is admitted through an anchor, subsequent service processing can be handled by the corresponding SFC domain.

Under this abstraction, the candidate pair $(j,k)$ jointly captures the selected access anchor and the selected bearer mode. Because service categories differ in access-orchestration delay budgets, reliability requirements, and data sizes, the controller must evaluate whether each anchor--bearer candidate can admit the request within its access-orchestration delay budget. A successful orchestration outcome means that the request is realized through a feasible anchor--bearer candidate within the access-orchestration stage, after which the corresponding SFC-capable service domain handles subsequent service processing. This definition keeps the model close to SFC orchestration while avoiding the excessive complexity of full-chain remapping in every short slot.

\subsection{TEMP-Assisted Rolling Access Orchestration}

TEMP denotes temporal elastic mapping and parking. It is introduced to represent a controlled temporal action for requests that are not served immediately but may still be realized in a future slot. Physically, TEMP can be supported by onboard storage, an aircraft-side storage-type VNF, or a lightweight buffering function managed by the SDN/NFV controller. Conceptually, TEMP is not a passive queue, but a schedulable temporal resource whose value depends on whether the request can be realized within its access-orchestration delay budget.

The TEMP mechanism is motivated by the time-varying nature of civil aviation SAGINs. A request that has no suitable immediate bearer may become feasible in a later slot because of changes in aircraft positions, satellite visibility, ground gateway availability, or resource contention. Therefore, discarding all currently infeasible requests may waste future communication opportunities. At the same time, admitting requests into TEMP without considering future realization may only postpone failures and increase buffer pressure. The controller therefore needs to determine which requests are worth preserving in TEMP according to realization probability, residence feasibility, and buffer pressure. The TEMP management state records carried-over requests, residence time, and buffer occupancy, which are used to determine whether a request can be further preserved or must be released through DROP.

At the beginning of each slot $t$, the controller observes the newly arrived requests and the carried-over TEMP requests. These requests form the active request set to be orchestrated in the current slot. For each active request, the controller selects exactly one action from NOW, TEMP, REJECT, and DROP. NOW consumes current aggregated data-link bearer capacity and realizes the request in the current access-orchestration slot if the access-orchestration delay budget is satisfied. TEMP consumes onboard temporal storage and preserves the request for possible future realization. REJECT applies to newly arrived requests that cannot be admitted, whereas DROP applies to carried-over TEMP requests that can no longer be legally preserved.

The resulting problem couples spatial bearer allocation and temporal preservation. Spatially, the controller allocates heterogeneous A2A, A2G, and A2S bearer resources to competing requests through feasible anchor--bearer candidates. Temporally, it determines whether currently unserved requests should be parked in TEMP according to look-ahead opportunities, residence limits, and buffer pressure. The finite-horizon future information used by TEMP is treated as an input to orchestration. In each rolling decision slot, the controller uses finite-horizon look-ahead information on future link opportunities; simulations use recorded future slots as look-ahead inputs to isolate the value of TEMP-aware orchestration, while dedicated traffic or link-state prediction is outside the scope of this paper.

The following section formulates this rolling online orchestration problem as a success-prioritized mathematical model. The model explicitly represents NOW realization, TEMP admission, request rejection, carried-over request dropping, data-link capacity, TEMP feasibility, buffer capacity, and lexicographic service objectives.
\section{Mathematical Model}
\label{sec:mathematical_model}

This section formulates the rolling-slot orchestration problem. At each decision slot, the controller assigns one action to every pending request. NOW, REJECT, and DROP take effect in the current slot, whereas TEMP preserves a request for possible future realization. The formulation therefore captures both current data-link bearer-resource allocation and deferred TEMP realization.

\subsection{Rolling Request State and Four-Action Semantics}
\label{subsec:request_state}

Let $t$ denote the current decision slot. The pending orchestration set consists of newly arrived requests and carried-over TEMP requests:
\begin{equation}
\mathcal{U}(t)
=
\mathcal{N}(t)\cup\mathcal{B}(t),
\label{eq:active_request_set}
\end{equation}
where $\mathcal{N}(t)$ is the set of newly arrived requests, $\mathcal{B}(t)$ is the set of carried-over TEMP requests, and
$\mathcal{N}(t)\cap\mathcal{B}(t)=\varnothing$.

Each active request $r\in\mathcal{U}(t)$ is characterized by
\begin{equation}
r
=
\left\langle
i(r),
\kappa(r),
s_r,
D_r^{\mathrm{acc}},
w_r(t)
\right\rangle,
\label{eq:request_tuple}
\end{equation}
where $i(r)$ is the associated aircraft, $\kappa(r)$ is the service category, $s_r$ is the data size, $D_r^{\mathrm{acc}}$ is the access-orchestration delay budget from request arrival to admission into an SFC-capable service domain, and $w_r(t)$ records the number of TEMP residence slots. For a newly arrived request $r\in\mathcal{N}(t)$,
\begin{equation}
w_r(t)=0.
\label{eq:new_waiting_time}
\end{equation}

Let $\mathcal{I}$ be the set of aircraft, $\mathcal{V}$ the set of candidate SFC access anchors, and $\mathcal{K}$ the set of aggregated data-link bearer modes. Following Section~\ref{sec:system_overview}, the ordered pair $(j,k)\in\mathcal{V}\times\mathcal{K}$ represents an anchor--bearer candidate, where $j$ indexes an SFC access anchor and $k$ indexes an aggregated data-link bearer mode.

For each request $r\in\mathcal{U}(t)$, the controller selects one action from NOW, TEMP, REJECT, and DROP. NOW assigns the request to immediate service, TEMP retains it for possible future realization, REJECT denotes non-admission of a newly arrived request, and DROP releases a carried-over TEMP request. Requests assigned NOW, REJECT, or DROP leave the pending set after the current decision, whereas requests assigned TEMP constitute $\mathcal{B}(t+1)$ with updated residence state $w_r(t+1)=w_r(t)+1$.

For each request $r\in\mathcal{U}(t)$, let
$\Gamma_r(t)\subseteq\mathcal{V}\times\mathcal{K}$
denote the set of anchor--bearer pairs that are currently reachable and service-compatible. Shared bearer-capacity competition is enforced later by the capacity constraints. If request $r$ is served through $(j,k)\in\Gamma_r(t)$ at slot $t$, its access-orchestration completion delay is
\begin{equation}
d^{\mathrm{now}}_{rjk}(t)
=
w_r(t)T_{\mathrm{slot}}
+
d^{\mathrm{q}}_{jk}(t)
+
\frac{s_r}{\mu_{jk}(t)}
+
d^{\mathrm{p}}_{i(r)j}(t),
\label{eq:now_delay}
\end{equation}
where $w_r(t)T_{\mathrm{slot}}$ is the accumulated TEMP residence delay, $d^{\mathrm{q}}_{jk}(t)$ is the queueing delay, $\mu_{jk}(t)$ is the available transmission rate, and $d^{\mathrm{p}}_{i(r)j}(t)$ is the propagation delay from aircraft $i(r)$ to access anchor $j$.

The delay-feasible NOW candidate set is
\begin{equation}
\mathcal{A}^{\mathrm{now}}_r(t)
=
\left\{
(j,k)\in\Gamma_r(t)
\;\middle|\;
d^{\mathrm{now}}_{rjk}(t)
\leq
D_r^{\mathrm{acc}}
\right\}.
\label{eq:now_action_set}
\end{equation}
Thus, NOW candidates satisfy reachability, service compatibility, and access-orchestration delay feasibility. REJECT is available only to newly arrived requests, whereas DROP is available only to carried-over TEMP requests; the corresponding request-type restrictions are imposed in Section~\ref{subsec:constraints}. Unlike REJECT, DROP releases a previously preserved TEMP request and represents an unsuccessful deferred-service outcome after temporal resources have already been occupied.

\subsection{Deferred-Realization Action: TEMP}
\label{subsec:temp_action}

TEMP is the deferred-realization action. A request assigned TEMP is retained at slot $t$ and may be realized in a future slot if a suitable anchor--bearer opportunity is predicted within its access-orchestration delay budget. Let $W_{\mathrm{pred}}$ be the finite look-ahead window:
\begin{equation}
\mathcal{T}^{\mathrm{pred}}(t)
=
\left\{
t+1,t+2,\ldots,t+W_{\mathrm{pred}}
\right\}.
\label{eq:pred_window}
\end{equation}

For each $\tau\in\mathcal{T}^{\mathrm{pred}}(t)$, let
\begin{equation}
\mathcal{A}^{\mathrm{fut}}_r(\tau\mid t)
\subseteq
\mathcal{V}\times\mathcal{K}
\label{eq:future_action_set}
\end{equation}
denote the set of anchor--bearer pairs predicted at slot $t$ to be reachable and service-compatible for request $r$ at slot $\tau$.

If request $r$ is retained in TEMP at slot $t$ and served at slot $\tau$ through $(j,k)$, its predicted access-orchestration completion delay is
\begin{equation}
\begin{aligned}
d^{\mathrm{fut}}_{rjk}(\tau\mid t)
={}&
w_r(t)T_{\mathrm{slot}}
+
(\tau-t)T_{\mathrm{slot}}
+
d^{\mathrm{q}}_{jk}(\tau\mid t)
\\
&+
\frac{s_r}{\mu_{jk}(\tau\mid t)}
+
d^{\mathrm{p}}_{i(r)j}(\tau\mid t).
\end{aligned}
\label{eq:future_delay}
\end{equation}
The first two terms represent accumulated residence and additional waiting delay, while the remaining terms are the predicted queueing, transmission, and propagation delays.

For each future candidate pair, let
\begin{equation}
\phi_{rjk}(\tau\mid t)\in[0,1],
\qquad
(j,k)\in\mathcal{A}^{\mathrm{fut}}_r(\tau\mid t),
\label{eq:future_realization_prob}
\end{equation}
denote the predicted conditional probability that request $r$ is served through $(j,k)$ at slot $\tau$, given that it has not been realized earlier. It is set to zero if the access-orchestration delay or residual-capacity condition is violated:
\begin{equation}
\phi_{rjk}(\tau\mid t)=0,
\quad
\text{if }
d^{\mathrm{fut}}_{rjk}(\tau\mid t)>D_r^{\mathrm{acc}}
\text{ or }
s_r>\bar{C}_{jk}(\tau\mid t),
\label{eq:future_prob_zero}
\end{equation}
where $\bar{C}_{jk}(\tau\mid t)$ is the residual capacity of data-link bearer resource $(j,k)$ at slot $\tau$ predicted at slot $t$. This parameter excludes individually infeasible future actions without reserving future capacity deterministically.

The single-slot realization probability, first-realization probability, and overall TEMP realization probability are respectively given by
\begin{align}
p_r(\tau\mid t)
&=
\max_{(j,k)\in\mathcal{A}^{\mathrm{fut}}_r(\tau\mid t)}
\phi_{rjk}(\tau\mid t),
\label{eq:single_slot_prob}
\\
\pi_r(\tau\mid t)
&=
p_r(\tau\mid t)
\prod_{\ell=t+1}^{\tau-1}
\left(
1-p_r(\ell\mid t)
\right),
\label{eq:first_realization_prob}
\\
P^{\mathrm{temp}}_r(t)
&=
\sum_{\tau\in\mathcal{T}^{\mathrm{pred}}(t)}
\pi_r(\tau\mid t),
\label{eq:temp_realization_prob}
\end{align}
with $p_r(\tau\mid t)=0$ if
$\mathcal{A}^{\mathrm{fut}}_r(\tau\mid t)=\emptyset$.

Define the positive-probability future action set and the feasible future-slot set as
\begin{align}
\mathcal{A}^{\mathrm{fut,feas}}_r(\tau\mid t)
&=
\left\{
(j,k)\in\mathcal{A}^{\mathrm{fut}}_r(\tau\mid t)
\;\middle|\;
\phi_{rjk}(\tau\mid t)>0
\right\},
\label{eq:future_feasible_action_set}
\\
\mathcal{T}^{\mathrm{feas}}_r(t)
&=
\left\{
\tau\in\mathcal{T}^{\mathrm{pred}}(t)
\;\middle|\;
\mathcal{A}^{\mathrm{fut,feas}}_r(\tau\mid t)\neq\emptyset
\right\}.
\label{eq:future_feasible_slot_set}
\end{align}
For each $\tau\in\mathcal{T}^{\mathrm{feas}}_r(t)$, the predicted delay associated with the maximum-probability pair, with minimum delay used to break ties, is
\begin{equation}
\bar{d}^{\mathrm{fut}}_r(\tau\mid t)
=
\min_{\substack{
(j,k)\in\mathcal{A}^{\mathrm{fut,feas}}_r(\tau\mid t)\\
\phi_{rjk}(\tau\mid t)=p_r(\tau\mid t)
}}
d^{\mathrm{fut}}_{rjk}(\tau\mid t).
\label{eq:min_future_delay}
\end{equation}

For $P^{\mathrm{temp}}_r(t)>0$, the conditional expected TEMP completion delay is
\begin{equation}
\bar{d}^{\mathrm{temp}}_r(t)
=
\frac{
\displaystyle
\sum_{\tau\in\mathcal{T}^{\mathrm{feas}}_r(t)}
\pi_r(\tau\mid t)
\bar{d}^{\mathrm{fut}}_r(\tau\mid t)
}{
P^{\mathrm{temp}}_r(t)
}.
\label{eq:temp_expected_delay}
\end{equation}
This quantity is conditioned on successful realization within the finite look-ahead window. If $P^{\mathrm{temp}}_r(t)=0$, TEMP is infeasible and the expected delay need not be evaluated. All look-ahead quantities above are precomputed inputs at slot $t$, rather than optimization variables.

\subsection{Decision Variables}
\label{subsec:decision_variables}

For each request $r\in\mathcal{U}(t)$, the four request-level actions are encoded by the following binary variables:
\begin{equation}
\begin{aligned}
x_{rjk}(t)&\in\{0,1\},
&& (j,k)\in\mathcal{A}^{\mathrm{now}}_r(t),\\
z_r(t),\ e_r(t),\ b^{\mathrm{drop}}_r(t)
&\in\{0,1\}.
\end{aligned}
\label{eq:decision_variables}
\end{equation}
Here, $x_{rjk}(t)$ indicates NOW assignment through $(j,k)$, $z_r(t)$ indicates TEMP preservation, $e_r(t)$ indicates REJECT, and $b^{\mathrm{drop}}_r(t)$ indicates DROP.

The auxiliary NOW-assignment indicator is
\begin{equation}
y_r(t)
=
\sum_{(j,k)\in\mathcal{A}^{\mathrm{now}}_r(t)}
x_{rjk}(t).
\label{eq:now_success_indicator}
\end{equation}
Under the action-uniqueness and feasibility constraints, $y_r(t)=1$ represents deterministic NOW realization, while TEMP contributes through $P^{\mathrm{temp}}_r(t)$.

\subsection{Feasibility Constraints}
\label{subsec:constraints}

For each request $r\in\mathcal{U}(t)$, exactly one action must be selected:
\begin{equation}
\sum_{(j,k)\in\mathcal{A}^{\mathrm{now}}_r(t)}
x_{rjk}(t)
+
z_r(t)
+
e_r(t)
+
b^{\mathrm{drop}}_r(t)
=
1.
\label{eq:action_uniqueness}
\end{equation}

REJECT and DROP follow request-type semantics:
\begin{align}
b^{\mathrm{drop}}_r(t)&=0,
&&\forall r\in\mathcal{N}(t),
\label{eq:new_no_drop}
\\
e_r(t)&=0,
&&\forall r\in\mathcal{B}(t).
\label{eq:old_no_reject}
\end{align}
Thus, newly arrived requests cannot be dropped, and carried-over TEMP requests cannot be rejected.

For each anchor--bearer resource $(j,k)$, the current-slot NOW capacity constraint is
\begin{equation}
\sum_{\substack{
r\in\mathcal{U}(t):\\
(j,k)\in\mathcal{A}^{\mathrm{now}}_r(t)
}}
s_r x_{rjk}(t)
\leq
C_{jk}(t),
\quad
\forall (j,k)\in\mathcal{V}\times\mathcal{K}.
\label{eq:current_capacity_constraint}
\end{equation}
REJECT and DROP consume no current data-link bearer capacity, while TEMP consumes buffer space and relies on future bearer opportunities.

TEMP is permitted only when the predicted realization probability and residence-time requirement are satisfied:
\begin{equation}
I^{\mathrm{temp}}_r(t)
=
\begin{cases}
1,
&
P^{\mathrm{temp}}_r(t)\geq P^{\mathrm{temp}}_{\min}
\ \text{and}\
w_r(t)+1\leq W^{\max}_r,\\
0,
&
\text{otherwise},
\end{cases}
\label{eq:temp_feasible_indicator}
\end{equation}
where $P^{\mathrm{temp}}_{\min}$ is the minimum acceptable TEMP realization probability and $W^{\max}_r$ is the maximum number of TEMP residence slots allowed for request $r$. The TEMP feasibility constraint is
\begin{equation}
z_r(t)\leq I^{\mathrm{temp}}_r(t),
\quad
\forall r\in\mathcal{U}(t).
\label{eq:temp_feasibility_constraint}
\end{equation}

For each aircraft $i$, the post-decision TEMP occupancy cannot exceed its TEMP buffer capacity:
\begin{equation}
\sum_{\substack{
r\in\mathcal{U}(t):\\
i(r)=i
}}
s_r z_r(t)
\leq
M_i,
\quad
\forall i\in\mathcal{I},
\label{eq:temp_buffer_constraint}
\end{equation}
where $M_i$ is the TEMP buffer capacity of aircraft $i$.

\subsection{Lexicographic Service Objective}
\label{subsec:lexicographic_objective}

The objective follows the operational priority of the rolling service process: expected service realization is optimized first, normalized access-orchestration delay second, and TEMP-related risk third:
\begin{equation}
\operatorname{lex\,max}
\left(
J_1(t),-J_2(t),-J_3(t)
\right).
\label{eq:lex_objective}
\end{equation}
The problem first maximizes $J_1(t)$, then minimizes $J_2(t)$ among the solutions attaining the optimal $J_1(t)$, and finally minimizes $J_3(t)$ without degrading the preceding objectives.

The first-layer objective maximizes expected service realization:
\begin{equation}
J_1(t)
=
\sum_{r\in\mathcal{U}(t)}
y_r(t)
+
\sum_{r\in\mathcal{U}(t)}
P^{\mathrm{temp}}_r(t)z_r(t).
\label{eq:j1_objective}
\end{equation}
The first term is deterministic NOW realization, whereas the second term is the probability-weighted TEMP realization benefit.

The second-layer objective minimizes expected normalized access-orchestration delay:
\begin{equation}
\begin{aligned}
J_2(t)
={}&
\sum_{r\in\mathcal{U}(t)}
\sum_{(j,k)\in\mathcal{A}^{\mathrm{now}}_r(t)}
\frac{d^{\mathrm{now}}_{rjk}(t)}
{D_r^{\mathrm{acc}}}
x_{rjk}(t)
\\
&+
\sum_{r\in\mathcal{U}(t)}
P^{\mathrm{temp}}_r(t)
\frac{\bar{d}^{\mathrm{temp}}_r(t)}
{D_r^{\mathrm{acc}}}
z_r(t).
\end{aligned}
\label{eq:j2_objective}
\end{equation}
Normalization makes requests with different access-orchestration delay budgets comparable. Since this layer is optimized only after $J_1(t)$ is fixed, delay reduction cannot sacrifice expected service realization.

The third-layer objective controls DROP and residual TEMP-related risk:
\begin{equation}
\begin{aligned}
J_3(t)
={}&
\chi_{\mathrm{drop}}
\sum_{r\in\mathcal{B}(t)}
b^{\mathrm{drop}}_r(t)
+
\omega_w
\sum_{r\in\mathcal{U}(t)}
z_r(t)
\frac{w_r(t)+1}{W^{\max}_r}
\\
&+
\omega_M
\sum_{i\in\mathcal{I}}
\frac{
\displaystyle
\sum_{\substack{
r\in\mathcal{U}(t)\\
i(r)=i
}}
s_r z_r(t)
}{
M_i
}.
\end{aligned}
\label{eq:j3_objective}
\end{equation}
The three terms penalize DROP, residence urgency, and post-decision TEMP buffer utilization, respectively. Low-realization-probability TEMP choices are already discouraged by $J_1(t)$ and the TEMP eligibility constraint, so $J_3(t)$ focuses on residual risks from DROP, prolonged residence, and buffer pressure.

The model jointly captures spatial data-link bearer opportunities provided by heterogeneous A2A, A2G, and A2S links and temporal elasticity provided by TEMP. Since all look-ahead quantities are precomputed parameters, the objective layers and constraints are linear in the binary decision variables, yielding a lexicographic MILP. Repeatedly solving this model over the full binary action space remains computationally expensive for rolling online control, which motivates the MRSAR algorithm developed next.
\section{MRSAR Algorithm}
\label{sec:mrsar_algorithm}

The lexicographic MILP in Section~\ref{sec:mathematical_model} defines the request-level action structure, feasibility conditions, and optimization priorities of the rolling orchestration problem. Directly solving the MILP at every slot becomes computationally expensive as the number of active requests and feasible anchor--bearer candidates increases. We therefore develop MRSAR (Model-Induced Risk and Scarcity-Aware Refinement), which constructs a feasible rolling-slot solution while reflecting the objective hierarchy and feasibility structure of the original model. MRSAR first ranks feasible NOW and TEMP candidates using model-induced risk and scarcity signals, and then improves the constructed solution through a bounded, feasibility-preserving neighborhood search.

The signals and moves used by MRSAR are derived from the objective and constraint structure of the MILP. The difference between deterministic NOW realization and probability-weighted TEMP realization in $J_1(t)$ induces deferred-realization risk, while the residence term in $J_3(t)$ provides the TEMP residence-urgency signal. The shared data-link bearer-capacity and TEMP buffer constraints provide resource-scarcity indicators, and the mutually exclusive request actions coupled through shared resources define the neighborhood moves. The TEMP realization probability $P_r^{\mathrm{temp}}(t)$ and conditional expected delay $\bar{d}_r^{\mathrm{temp}}(t)$ are precomputed inputs to MRSAR.

\subsection{Model-Induced Request Risk and Resource Scarcity}
\label{subsec:risk_scarcity}

According to the first-layer objective in \eqref{eq:j1_objective}, NOW contributes a deterministic realization benefit of one, whereas TEMP contributes the probability-weighted benefit $P^{\mathrm{temp}}_r(t)$. The deferred-realization risk of request $r$ is therefore defined as
\begin{equation}
g_r(t)
=
1-P^{\mathrm{temp}}_r(t).
\label{eq:deferred_risk}
\end{equation}
A larger $g_r(t)$ indicates a greater potential loss in the first-layer objective if the request is deferred instead of being served through NOW.

The residence urgency after preserving request $r$ for one additional slot is
\begin{equation}
h_r(t)
=
\frac{w_r(t)+1}{W^{\max}_r}.
\label{eq:residence_urgency}
\end{equation}
Combining deferred-realization risk and residence urgency gives
\begin{equation}
\xi_r(t)
=
g_r(t)
+
\alpha_h h_r(t),
\label{eq:request_risk_measure}
\end{equation}
where $\alpha_h\geq0$ controls the influence of residence urgency. A larger $\xi_r(t)$ indicates that further postponement is less desirable.

For an anchor--bearer resource $(j,k)$ with $C_{jk}(t)>0$, its current-slot scarcity indicator is
\begin{equation}
\lambda^{\mathrm{now}}_{jk}(t)
=
1+
\alpha_C
\frac{
\displaystyle
\sum_{r\in\mathcal{U}(t)}
s_r
\mathbf{1}
\left\{
(j,k)\in\mathcal{A}^{\mathrm{now}}_r(t)
\right\}
}{
C_{jk}(t)
},
\label{eq:now_scarcity}
\end{equation}
where $\alpha_C\geq0$ controls the sensitivity to data-link bearer-resource pressure. The indicator increases when a resource is shared by more candidate requests or provides less available capacity.

The TEMP buffer scarcity of aircraft $i$ is defined as
\begin{equation}
\lambda^{\mathrm{temp}}_i(t)
=
1+
\alpha_M
\frac{
\displaystyle
\sum_{\substack{
r\in\mathcal{U}(t)\\
i(r)=i
}}
s_r I^{\mathrm{temp}}_r(t)
}{
M_i
},
\label{eq:temp_scarcity}
\end{equation}
where $\alpha_M\geq0$ controls the sensitivity to TEMP buffer pressure. Both scarcity indicators are computed once at the beginning of slot $t$; residual data-link bearer and buffer capacities are then updated during construction and refinement. Request risk determines service urgency, whereas resource scarcity reflects the opportunity cost of NOW or TEMP actions.

\subsection{Candidate Action Valuation and Feasible Solution Construction}
\label{subsec:action_valuation}

The request-risk and resource-scarcity measures are translated into valuation criteria for candidate NOW and TEMP actions. For each feasible NOW candidate $(r,j,k)$, define
\begin{equation}
\operatorname{Val}^{\mathrm{now}}_{rjk}(t)
=
\frac{
\xi_r(t)
}{
\lambda^{\mathrm{now}}_{jk}(t)s_r
}
-
\alpha_d
\frac{
d^{\mathrm{now}}_{rjk}(t)
}{
D_r^{\mathrm{acc}}
},
\label{eq:now_valuation}
\end{equation}
where $\alpha_d\geq0$ is the NOW delay coefficient. The first term measures the postponement risk reduced per unit of scarce data-link bearer resource, whereas the second term accounts for normalized access-orchestration delay.

For each TEMP-feasible request, define
\begin{equation}
\begin{aligned}
\operatorname{Val}^{\mathrm{temp}}_r(t)
={}&
\frac{
P^{\mathrm{temp}}_r(t)
}{
\lambda^{\mathrm{temp}}_{i(r)}(t)s_r
}
-
\beta_d
\frac{
\bar{d}^{\mathrm{temp}}_r(t)
}{
D_r^{\mathrm{acc}}
}
\\
&-
\beta_h h_r(t),
\end{aligned}
\label{eq:temp_valuation}
\end{equation}
where $\beta_d,\beta_h\geq0$ are the TEMP delay and residence-urgency coefficients. A high TEMP valuation corresponds to promising future realization, limited buffer occupation, moderate expected delay, and sufficient remaining residence time.

The candidate valuations do not replace the lexicographic objective or feasibility constraints. They provide efficient estimates of the relative merits of candidate assignments by accounting for request urgency and resource-induced opportunity cost. Every assignment is still accepted only after delay, capacity, residence, buffer, and request-type constraints have been verified.

MRSAR first initializes the residual data-link bearer capacities and TEMP buffer capacities. Feasible NOW candidates are ranked in descending order of $\operatorname{Val}^{\mathrm{now}}_{rjk}(t)$, with smaller $s_r$ used to break ties. A candidate is accepted if the request has not been assigned and the corresponding residual bearer capacity is sufficient; then the residual capacity of $(j,k)$ is updated and the remaining NOW candidates of request $r$ are removed.

The remaining TEMP-feasible requests are ranked according to $\operatorname{Val}^{\mathrm{temp}}_r(t)$, again with smaller $s_r$ used to break ties. A request is retained in TEMP if the corresponding aircraft has sufficient residual buffer capacity. Any remaining request in $\mathcal{N}(t)$ is assigned REJECT, whereas any remaining request in $\mathcal{B}(t)$ is assigned DROP.

NOW is considered before TEMP during initialization because it provides deterministic realization, whereas TEMP provides probability-weighted future realization. This NOW-first rule is only an initialization strategy; later refinement may replace a NOW assignment if the resulting local modification improves the hierarchy-weighted objective gain. The construction therefore yields a complete feasible solution $\mathcal{S}^{(0)}(t)$ without enumerating the full binary action space.

\subsection{Model-Structured Neighborhood Refinement}
\label{subsec:model_structured_refinement}

The refinement neighborhood is constructed according to the block structure of the decision variables and their coupling through shared resource constraints. For each request, the NOW variables and the TEMP, REJECT, and DROP variables form a mutually exclusive action block. The data-link bearer-capacity and TEMP buffer constraints further couple different request blocks, motivating coordinated adjustments.

Let $\mathcal{S}^{(\mathrm{iter})}(t)$ denote the solution after refinement iteration $\mathrm{iter}$. MRSAR constructs a bounded neighborhood $\mathcal{M}(\mathcal{S}^{(\mathrm{iter})}(t))$ using three move types.

\begin{itemize}
    \item \emph{Rescue}: change a request currently assigned TEMP, REJECT, or DROP to a feasible NOW action, or change a request currently assigned REJECT or DROP to a feasible TEMP action.

    \item \emph{Rebind}: replace the current anchor--bearer pair of a NOW request with another pair in $\mathcal{A}^{\mathrm{now}}_r(t)$.

    \item \emph{Exchange}: jointly modify two request assignments to resolve bearer or TEMP resource competition that cannot be improved by a single-request move.
\end{itemize}

Only moves satisfying all model constraints are retained in the feasible neighborhood $\mathcal{M}^{\mathrm{feas}}(\mathcal{S}^{(\mathrm{iter})}(t))$. For a feasible move $m$, let
\begin{equation}
\mathcal{S}'(t)
=
\operatorname{Apply}
\left(
\mathcal{S}^{(\mathrm{iter})}(t),m
\right)
\label{eq:apply_move}
\end{equation}
be the solution obtained after applying $m$. The change in objective layer $\ell$ is
\begin{equation}
\Delta J_\ell(m;t)
=
J_\ell\bigl(\mathcal{S}'(t)\bigr)
-
J_\ell\bigl(\mathcal{S}^{(\mathrm{iter})}(t)\bigr),
\quad
\ell=1,2,3.
\label{eq:objective_change}
\end{equation}

To preserve the intended objective hierarchy during local search, the move gain is
\begin{equation}
\Psi(m;t)
=
\Omega_1\Delta J_1(m;t)
-
\Omega_2\Delta J_2(m;t)
-
\Delta J_3(m;t),
\label{eq:move_gain}
\end{equation}
where $\Omega_1\gg\Omega_2\gg1$ are fixed hierarchy weights chosen to approximate the lexicographic priority. The first term rewards service-realization improvement, whereas the second and third terms penalize increases in normalized delay and DROP/TEMP-related risk, respectively.

At iteration $\mathrm{iter}$, MRSAR selects
\begin{equation}
m^\star
=
\arg\max_{
m\in
\mathcal{M}^{\mathrm{feas}}
(\mathcal{S}^{(\mathrm{iter})}(t))
}
\Psi(m;t).
\label{eq:best_refinement_move}
\end{equation}
If $\Psi(m^\star;t)>0$, the solution is updated as
\begin{equation}
\mathcal{S}^{(\mathrm{iter}+1)}(t)
=
\operatorname{Apply}
\left(
\mathcal{S}^{(\mathrm{iter})}(t),m^\star
\right);
\label{eq:refinement_update}
\end{equation}
otherwise, the refinement process terminates.

The construction checks the original feasibility conditions before each assignment, and refinement evaluates only feasible Rescue, Rebind, and Exchange moves. Hence every accepted solution remains feasible. Since each accepted move has a strictly positive hierarchy-weighted gain and the number of iterations is bounded by $Q_{\max}$, the algorithm terminates with a locally improved feasible solution within the examined bounded neighborhood.

\subsection{Overall MRSAR Procedure}
\label{subsec:mrsar_procedure}

The complete online procedure is summarized in Algorithm~\ref{alg:mrsar}. The TEMP realization quantities are provided as precomputed inputs, while request risk, static resource scarcity, candidate valuations, feasible construction, and neighborhood refinement are performed online. The initial feasible solution follows the valuation-based construction in Section~\ref{subsec:action_valuation}, including the NOW-first rule, smaller-data-volume tie breaker, and explicit feasibility checking.

\begin{algorithm}[t]
\caption{MRSAR for Online Aggregated Data-Link Orchestration}
\label{alg:mrsar}
\begin{algorithmic}[1]
\Require Active request set $\mathcal{U}(t)$, NOW candidate sets
$\mathcal{A}^{\mathrm{now}}_r(t)$, precomputed TEMP quantities
$P^{\mathrm{temp}}_r(t)$ and $\bar d^{\mathrm{temp}}_r(t)$,
capacities $C_{jk}(t)$ and $M_i$, model and algorithm parameters,
and iteration limit $Q_{\max}$
\Ensure Decisions $x_{rjk}(t)$, $z_r(t)$, $e_r(t)$, and
$b^{\mathrm{drop}}_r(t)$

\State Compute $I^{\mathrm{temp}}_r(t)$, request-risk measures,
resource-scarcity indicators, and candidate valuations
\State Construct the initial feasible solution
$\mathcal{S}^{(0)}(t)$ using the NOW-first valuation rule

\For{$\mathrm{iter}=0$ to $Q_{\max}-1$}
    \State Generate feasible Rescue, Rebind, and Exchange moves
    $\mathcal{M}^{\mathrm{feas}}(\mathcal{S}^{(\mathrm{iter})}(t))$
    \If{$\mathcal{M}^{\mathrm{feas}}(\mathcal{S}^{(\mathrm{iter})}(t))
    =\emptyset$}
        \State \textbf{break}
    \EndIf

    \State $m^\star \gets
    \arg\max_{m\in\mathcal{M}^{\mathrm{feas}}
    (\mathcal{S}^{(\mathrm{iter})}(t))}
    \Psi(m;t)$

    \If{$\Psi(m^\star;t)\leq 0$}
        \State \textbf{break}
    \EndIf

    \State $\mathcal{S}^{(\mathrm{iter}+1)}(t)
    \gets
    \operatorname{Apply}
    (\mathcal{S}^{(\mathrm{iter})}(t),m^\star)$
\EndFor

\State Extract $x_{rjk}(t)$, $z_r(t)$, $e_r(t)$, and
$b^{\mathrm{drop}}_r(t)$ from the final solution
\end{algorithmic}
\end{algorithm}

\subsection{Online Complexity Analysis}
\label{subsec:online_complexity}

Let
\begin{equation}
n_{\mathrm{act}}(t)
=
\left|
\mathcal{U}(t)
\right|
\label{eq:active_request_number}
\end{equation}
be the number of active requests at slot $t$. Let $a^{\max}_{\mathrm{now}}$ be the maximum number of NOW candidates retained for each request, $a^{\max}_{\mathrm{mv}}$ the maximum number of neighborhood moves evaluated in each refinement iteration, and $c_{\mathrm{loc}}$ the incremental cost of checking the feasibility and objective change of one local move.

The request-risk measures, static scarcity indicators, and candidate valuations require at most
\begin{equation}
O\!\left(
n_{\mathrm{act}}(t)
a^{\max}_{\mathrm{now}}
\right)
\end{equation}
operations. Sorting the NOW candidates dominates the initial construction cost and requires
\begin{equation}
O\!\left(
n_{\mathrm{act}}(t)
a^{\max}_{\mathrm{now}}
\log
\left(
n_{\mathrm{act}}(t)
a^{\max}_{\mathrm{now}}
\right)
\right).
\label{eq:construction_complexity}
\end{equation}
The TEMP candidate sorting cost is at most $O(n_{\mathrm{act}}(t)\log n_{\mathrm{act}}(t))$ and is covered by the preceding term.

Each refinement iteration evaluates at most $a^{\max}_{\mathrm{mv}}$ feasible moves. By maintaining residual capacities, buffer occupancies, current actions, and objective components, the feasibility and gain of a move can be evaluated incrementally. The refinement cost is therefore
\begin{equation}
O\!\left(
Q_{\max}
a^{\max}_{\mathrm{mv}}
c_{\mathrm{loc}}
\right).
\label{eq:refinement_complexity}
\end{equation}

Accordingly, the overall online complexity of MRSAR is
\begin{equation}
\begin{aligned}
O\Bigl(
&
n_{\mathrm{act}}(t)
a^{\max}_{\mathrm{now}}
\log
\bigl(
n_{\mathrm{act}}(t)
a^{\max}_{\mathrm{now}}
\bigr)
\\
&+
Q_{\max}
a^{\max}_{\mathrm{mv}}
c_{\mathrm{loc}}
\Bigr).
\end{aligned}
\label{eq:mrsar_online_complexity}
\end{equation}
This analysis excludes the computation of TEMP look-ahead quantities, which are precomputed inputs. MRSAR therefore replaces repeated full MILP solving in online control with candidate valuation, feasibility-preserving construction, and bounded local refinement.
\section{Simulation Evaluation}
\label{sec:simulation}

This section evaluates the proposed MRSAR algorithm in a rolling online orchestration scenario. The evaluation follows the primary logic of the proposed formulation: The evaluation first examines service success and delay, then analyzes failure composition and TEMP utilization, followed by near-optimality, runtime scalability, ablation results, and look-ahead window sensitivity.

\subsection{Simulation Setup}
\label{subsec:simulation_setup}

We develop a trace-driven discrete-event simulation for a
high-density trans-Atlantic civil-aviation SAGIN. The air
segment is instantiated using real OAG flight schedules
involving 513 active commercial aircraft, while the space
segment uses AGI STK ephemerides for 66 Iridium satellites.
In addition, 20 ground gateways are deployed on both sides
of the Atlantic. At each slot, the A2A, A2G, and A2S
adjacency matrices are regenerated from the time-varying node
states according to geometric visibility, line-of-sight (LoS),
coverage, and link-budget constraints. The resulting slot-wise
adjacency matrices determine the reachable anchor--bearer
candidates and their current and look-ahead availability states.
Therefore, the evaluated dynamic topology is driven by real
flight-schedule data and physical connectivity conditions,
rather than by independently generated random links.

The considered data-link technologies include low-rate
safety-oriented cockpit links, such as VDL-Mode-2, HFDL,
legacy SATCOM, and SBB-Safety, as well as broadband
cabin-oriented links, such as ATG, 5G-ATG, and LEO broadband. The traffic contains both delay-sensitive cockpit services and bandwidth-intensive cabin services, including VoIP-Safety, CPDLC, ADS-B/C, AOC-Legacy, passenger VoIP, Web, and video sessions.

The simulation lasts for 500 slots. The slot duration is set to 0.2 s to emulate the fine-grained evolution of traffic arrivals, link availability, and TEMP residence states. The first 20 slots are used as the warm-up period and are excluded from the statistics. Since TEMP requests may be realized in later slots, all success, delay, and failure metrics are evaluated by arrival windows with a window length of 10 slots. A request is counted as successful only if it is eventually delivered within its QoS deadline. The traffic load factor is varied in ${0.6,0.8,1.0,1.2,1.4}$ by scaling the request data-rate demands while keeping the network configuration unchanged.

All runtime results are measured on the same general-purpose personal computer using an unoptimized research prototype, and are used for relative computational comparison among different solvers.

\subsection{Baselines and Metrics}
\label{subsec:baseline_metrics}

MRSAR is compared with one MILP-based reference and two greedy baselines. Gurobi-MILP directly solves the original mixed-integer formulation and serves as the optimal or near-optimal reference. When the MILP solver reaches the prescribed time limit in large-scale cases, the best feasible solution returned by Gurobi is recorded. First-Come-First-Served Greedy (FCFS-Greedy) processes requests according to their arrival order and selects a feasible immediate action whenever possible. Minimum-Delay Greedy (MinDelay-Greedy) ranks feasible NOW actions according to their normalized immediate delay and greedily occupies the currently shortest-delay resources. Both greedy baselines admit requests into TEMP only when the TEMP feasibility condition is satisfied.

The main metrics include the actual QoS success ratio, average normalized delay, total failure ratio with REJECT/DROP decomposition, QoS success-rate gap to the Gurobi-MILP reference, online solver time, and speedup over Gurobi-MILP. The success ratio is computed over arrival windows and counts only requests eventually delivered within their deadlines. The normalized delay is averaged over successful requests and measures the fraction of the delay budget consumed by service completion. The runtime is measured under the same single-machine research prototype for all compared methods.

\subsection{Overall QoS Success and Delay Performance}
\label{subsec:overall_performance}

% Unified size for simulation figures
\newcommand{\simfigwidth}{0.86\columnwidth}
\newcommand{\simfigheight}{0.24\textheight}

\begin{figure}[!t]
\centering
\includegraphics[
width=\simfigwidth,
height=\simfigheight,
keepaspectratio
]{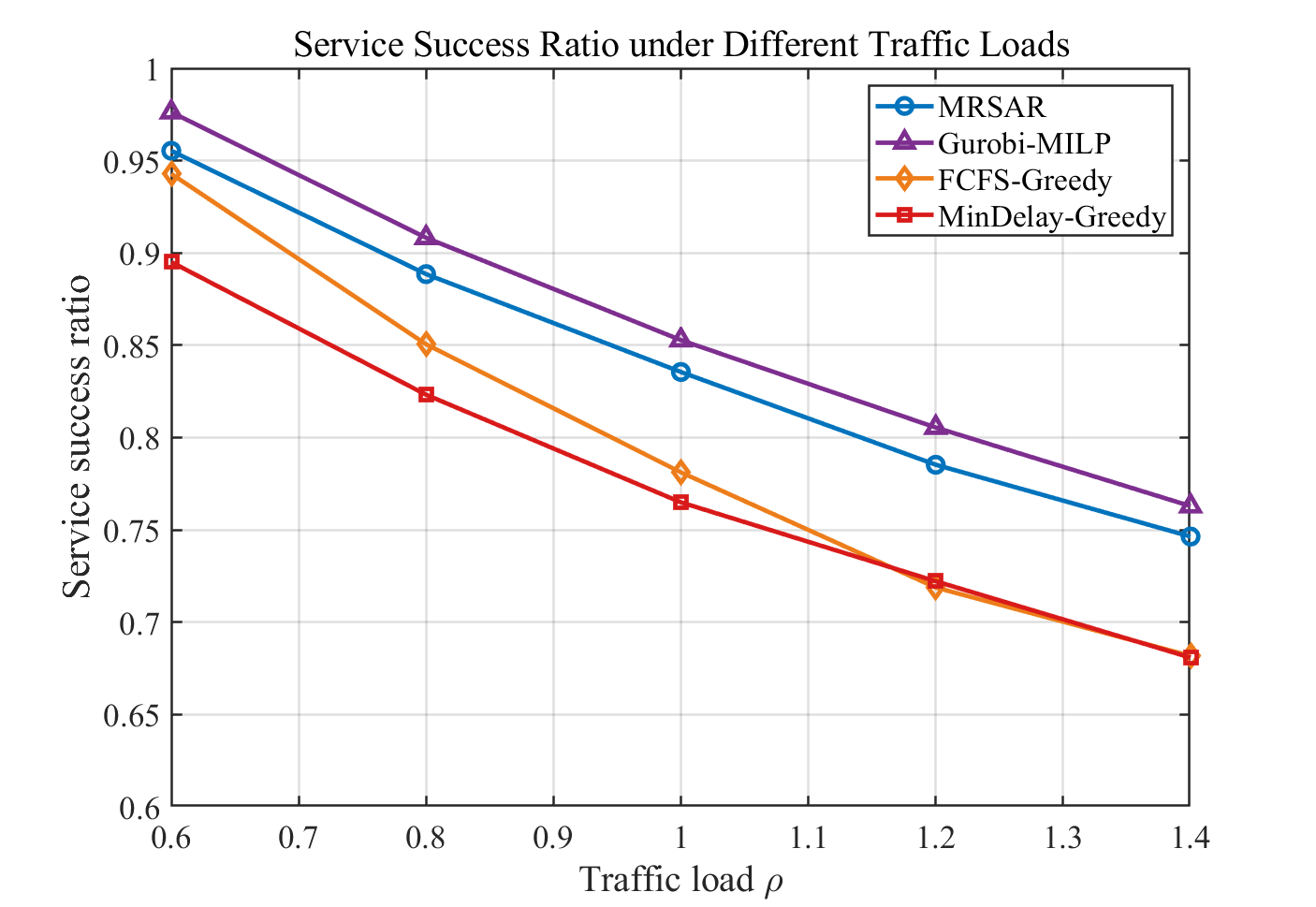}
\caption{Actual QoS success ratio under different traffic loads.}
\label{fig:success_vs_load}
\end{figure}

Fig.~\ref{fig:success_vs_load} reports the actual QoS success ratio under different traffic loads. As the load increases from 0.6 to 1.4, the success ratio of all algorithms decreases because more requests compete for limited A2A, A2G, and A2S data-link opportunities. MRSAR remains close to the Gurobi-MILP reference over the whole load range and consistently outperforms FCFS-Greedy and MinDelay-Greedy. The advantage becomes more evident under medium and high loads, where purely arrival-order-driven or delay-myopic decisions are more likely to consume scarce resources inefficiently. This confirms that the model-induced risk and scarcity signals in MRSAR are beneficial for the success-prioritized orchestration objective.

\begin{figure}[!t]
\centering
\includegraphics[
width=\simfigwidth,
height=\simfigheight,
keepaspectratio
]{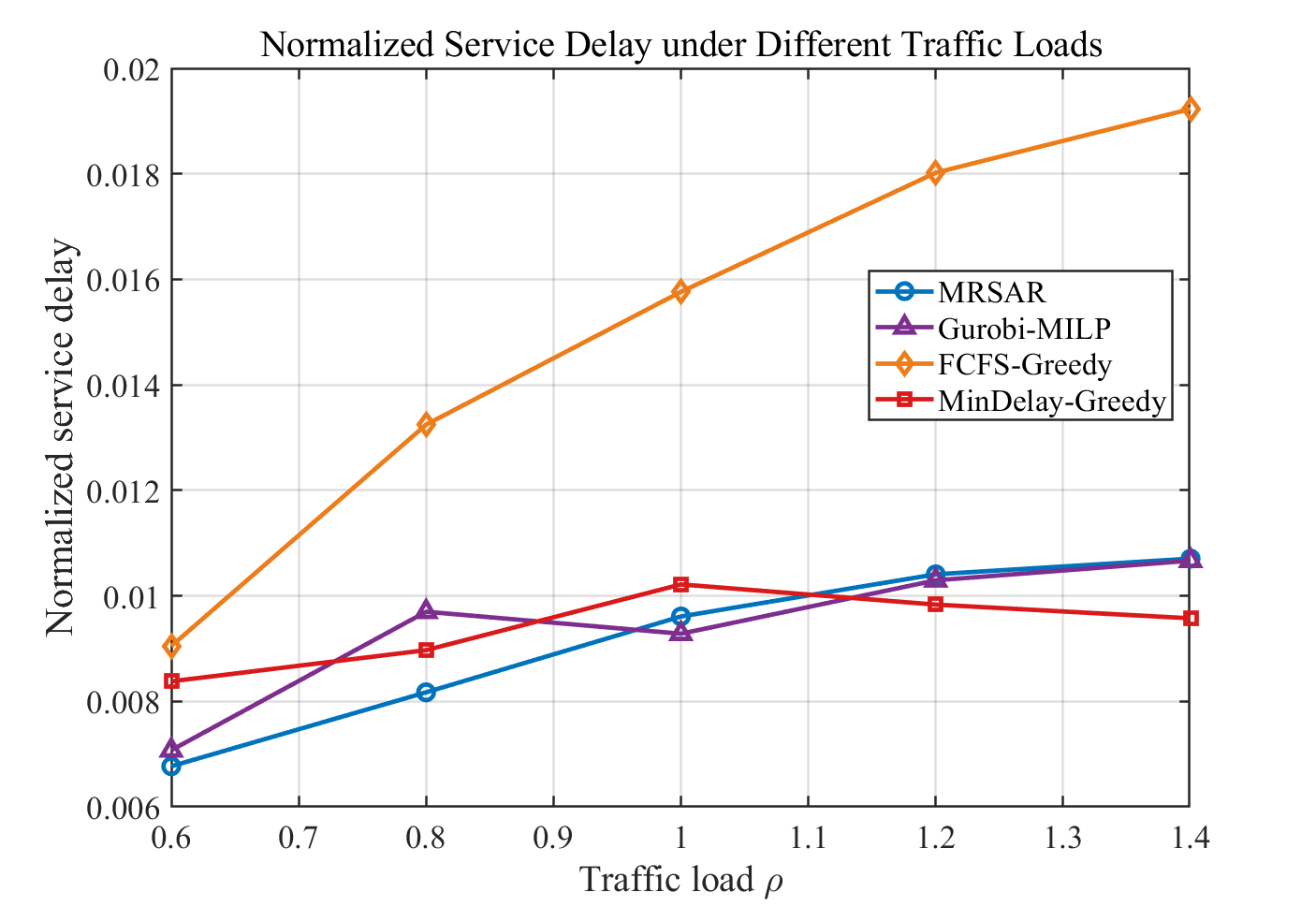}
\caption{Average normalized delay under different traffic loads.}
\label{fig:normalized_delay}
\end{figure}

Fig.~\ref{fig:normalized_delay} compares the average normalized service delay under different traffic loads. As the traffic load increases, FCFS-Greedy exhibits the most obvious delay growth, rising from about 0.009 at $\rho=0.6$ to nearly 0.019 at $\rho=1.4$. This is because arrival-order admission ignores delay-budget efficiency and may occupy favorable bearer resources with requests that are not delay-efficient. In contrast, MRSAR keeps the normalized delay within a low range, increasing moderately from about 0.007 to about 0.011 as the load becomes heavier. Its delay curve remains close to the Gurobi-MILP reference over the whole load range, which indicates that the proposed refinement mechanism improves the QoS success ratio without causing a significant delay penalty.

MinDelay-Greedy achieves competitive or even lower delay under some medium and high load settings. However, this does not imply better overall orchestration performance, since the delay metric is averaged only over successfully completed requests. A delay-myopic policy tends to favor requests with short immediate service delay and may reject or postpone requests that are more valuable for the success-prioritized objective. Therefore, its lower delay is partly obtained by serving an easier successful request set. Overall, Fig.~\ref{fig:normalized_delay} shows that MRSAR maintains delay performance comparable to Gurobi-MILP while preserving the primary advantage in QoS success under increasing traffic loads.

\subsection{Failure Decomposition and TEMP-Assisted Orchestration}

\begin{figure}[!t]
\centering
\includegraphics[
width=\simfigwidth,
height=\simfigheight,
keepaspectratio
]{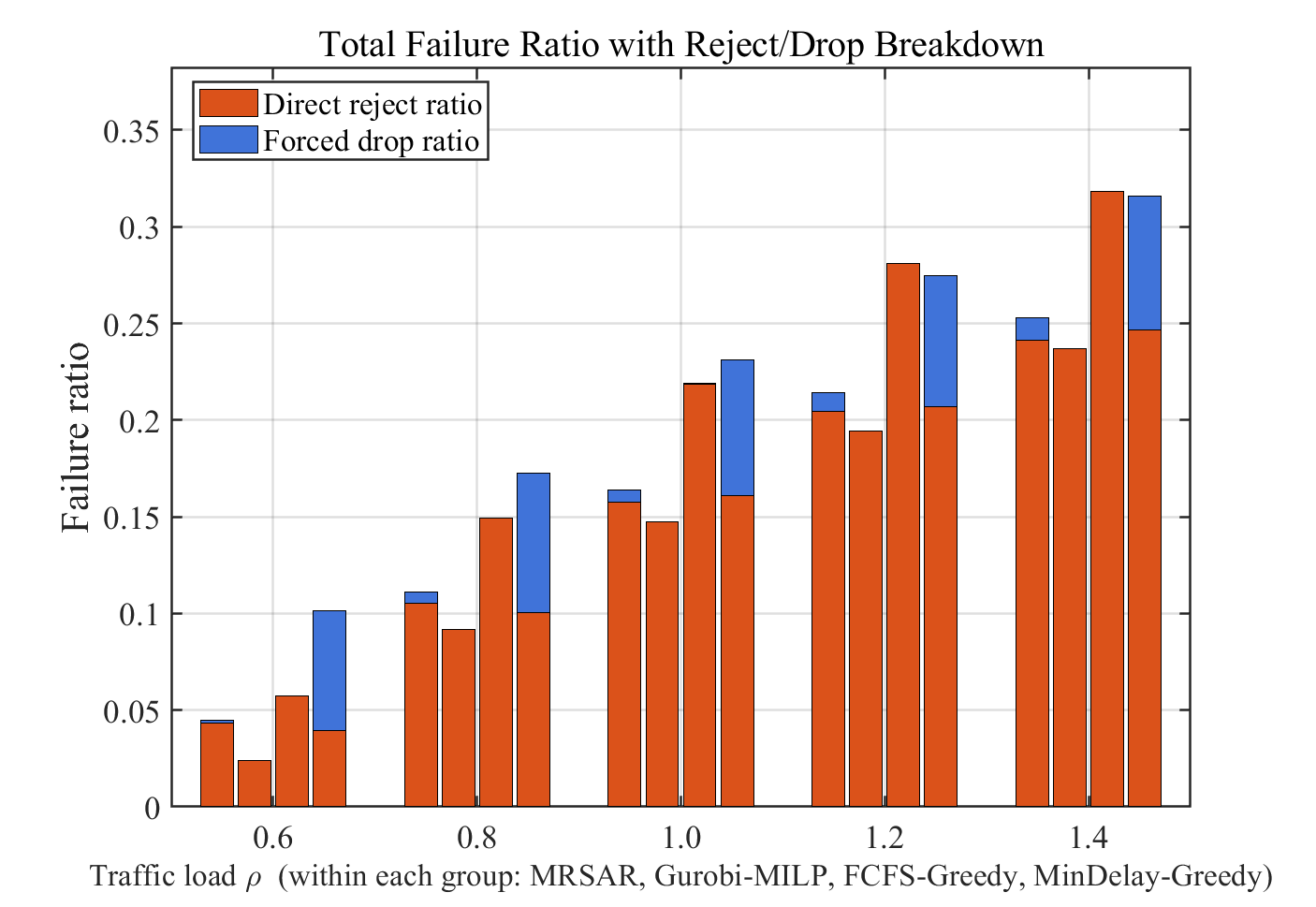}
\caption{Failure ratio with REJECT/DROP decomposition under different traffic loads. In each load group, the algorithms are ordered as MRSAR, Gurobi-MILP, FCFS-Greedy, and MinDelay-Greedy.}
\label{fig:failure_breakdown}
\end{figure}

Fig.~\ref{fig:failure_breakdown} decomposes the total failure ratio into direct REJECT and forced DROP components. The total failure ratio increases with the traffic load for all algorithms, which is consistent with the intensified data-link contention observed in Fig.~\ref{fig:success_vs_load}. MRSAR maintains a failure level close to the Gurobi-MILP reference and lower than the two greedy baselines. FCFS-Greedy mainly suffers from direct rejection because early arrival-order decisions may consume resources before later valuable requests are considered. MinDelay-Greedy may generate a more visible DROP component, indicating that short-delay admission does not necessarily guarantee reliable future realization. MRSAR keeps the DROP component controlled while reducing the overall failure ratio, which shows that TEMP in MRSAR is used as a managed temporal resource rather than a mechanism that merely postpones failures.

\label{subsec:temp_failure}

\begin{figure}[!t]
\centering
\includegraphics[
width=\simfigwidth,
height=\simfigheight,
keepaspectratio
]{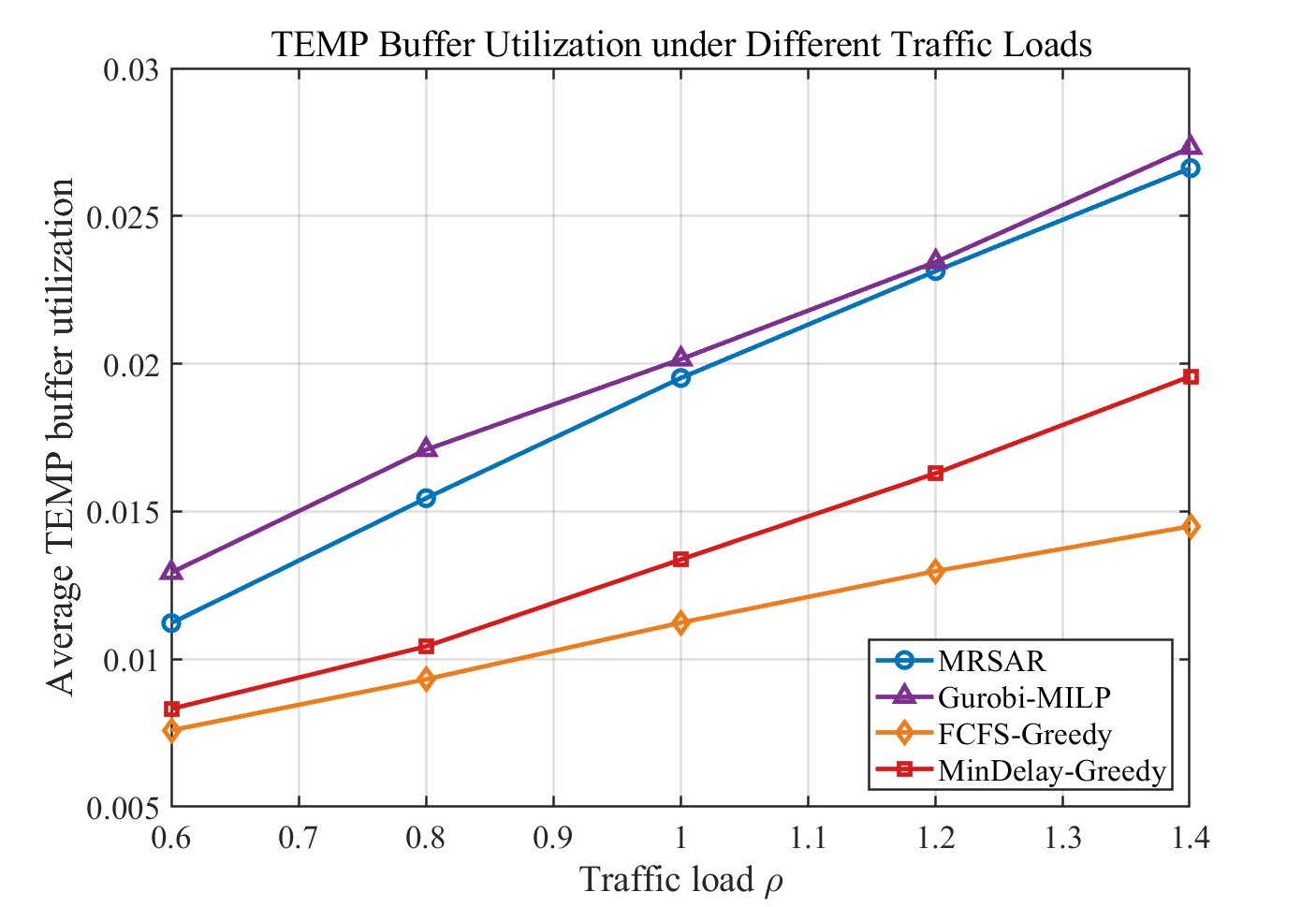}
\caption{Average TEMP buffer utilization under different traffic loads.}
\label{fig:temp_buffer_utilization}
\end{figure}

After the failure behavior is examined in Fig.~\ref{fig:failure_breakdown}, Fig.~\ref{fig:temp_buffer_utilization} further reveals the underlying TEMP-assisted mechanism. As the traffic load increases from $\rho=0.6$ to $\rho=1.4$, the TEMP buffer utilization of all algorithms increases, indicating that heavier traffic makes temporal parking more necessary. However, the utilization levels are clearly different. FCFS-Greedy maintains the lowest TEMP utilization over the whole load range, increasing only from about 0.008 to about 0.015. MinDelay-Greedy uses TEMP slightly more, but its utilization remains much lower than that of MRSAR and Gurobi-MILP. In contrast, MRSAR keeps a consistently high TEMP utilization and follows a trend close to the Gurobi-MILP reference, rising from about 0.011 at $\rho=0.6$ to nearly 0.027 at $\rho=1.4$. This shows that MRSAR can actively exploit temporal service opportunities instead of relying mainly on immediate admission.

This observation should be interpreted together with the failure decomposition in Fig.~\ref{fig:failure_breakdown}. A higher TEMP utilization is beneficial only when the preserved requests can be effectively realized later; otherwise, TEMP would merely postpone failures and increase the forced DROP ratio. As shown in Fig.~\ref{fig:failure_breakdown}, MRSAR maintains a failure level close to Gurobi-MILP and keeps the DROP component controlled under increasing traffic loads. Therefore, the high TEMP utilization of MRSAR does not come from uncontrolled buffering. Instead, it reflects effective deferred realization: requests are preserved only when they have sufficient future realization opportunities and acceptable resource impact. This explains an important source of the success-ratio improvement of MRSAR. Although all compared algorithms are allowed to use TEMP, MRSAR uses it more effectively by combining model-induced request risk values with scarcity-aware resource signals, thereby selecting which requests should consume immediate resources, which should be temporally parked, and which should be released.

\subsection{Near-Optimality and Runtime Scalability}
\label{subsec:gap_runtime}

\begin{figure}[!t]
\centering
\includegraphics[
width=\simfigwidth,
height=\simfigheight,
keepaspectratio
]{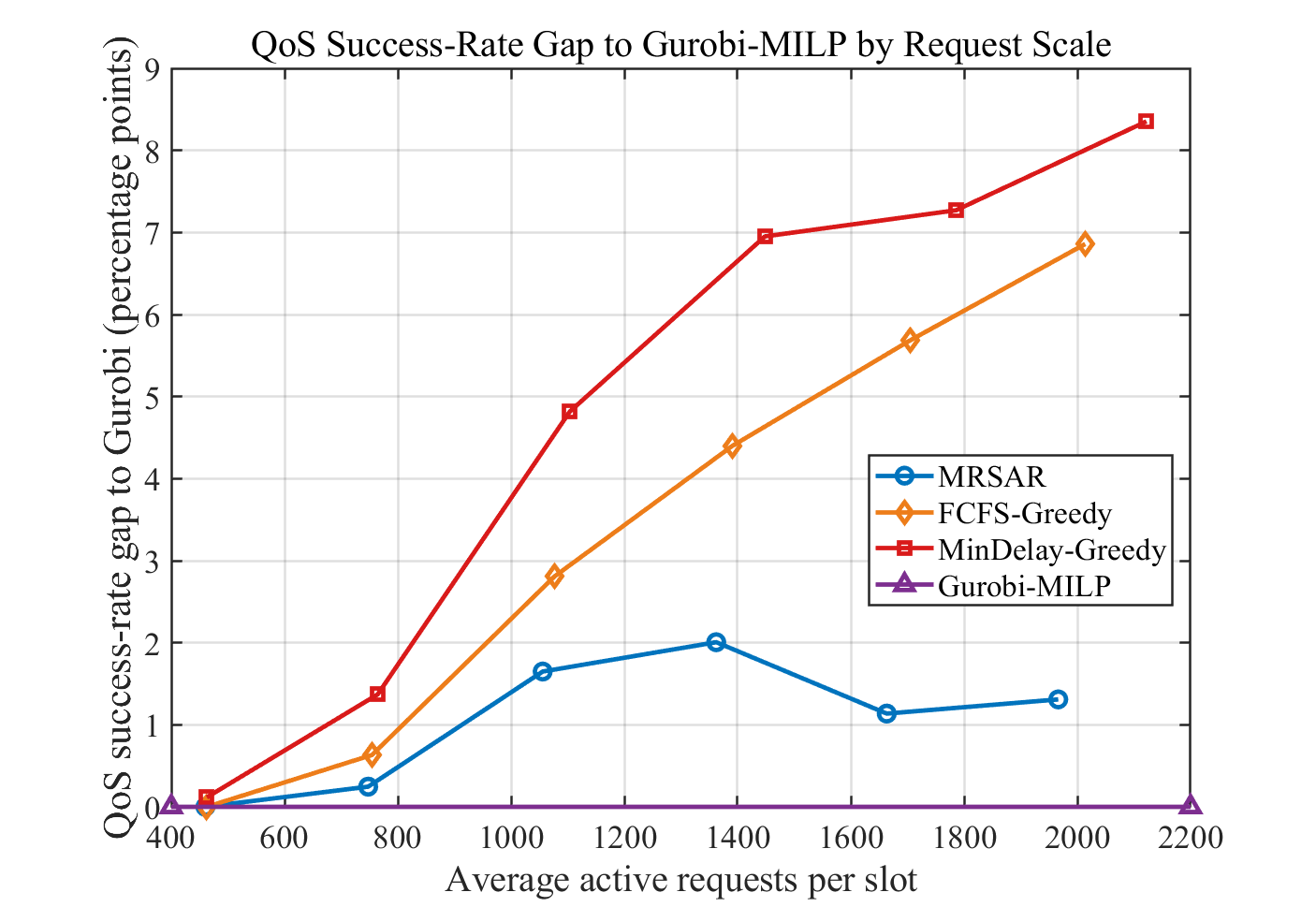}
\caption{QoS success-rate gap to the Gurobi-MILP reference under different request scales.}
\label{fig:gap_gurobi}
\end{figure}

Fig.~\ref{fig:gap_gurobi} reports the QoS success-rate gap to the Gurobi-MILP reference under different request scales. The Gurobi-MILP curve is zero by definition, while the gap of each heuristic method measures how many percentage points of QoS success ratio are lost compared with the MILP reference. As the average number of active requests per slot increases, the gaps of the two greedy baselines grow rapidly. FCFS-Greedy increases from nearly zero at the light-load case to about 6.8 percentage points at large request scales, and MinDelay-Greedy rises even more sharply, reaching more than 8 percentage points. This indicates that arrival-order admission and delay-myopic selection become increasingly insufficient when the competition among data-link resources becomes stronger.

In contrast, MRSAR keeps the gap consistently small over the whole request-scale range. Its gap remains below about 2 percentage points, with a moderate peak around the medium-scale region and then decreases to nearly 1--1.3 percentage points in larger cases. This non-monotonic behavior is reasonable because the success gap is affected not only by the number of requests, but also by the feasible-action structure, TEMP realization opportunities, and resource contention pattern in each scale setting. The bounded gap confirms that MRSAR preserves most of the success-rate advantage of Gurobi-MILP while avoiding full MILP search. Together with the runtime results, this shows that MRSAR provides a more effective tradeoff between solution quality and online computational cost than the two greedy baselines.

\begin{figure}[!t]
\centering
\includegraphics[
width=\simfigwidth,
height=\simfigheight,
keepaspectratio
]{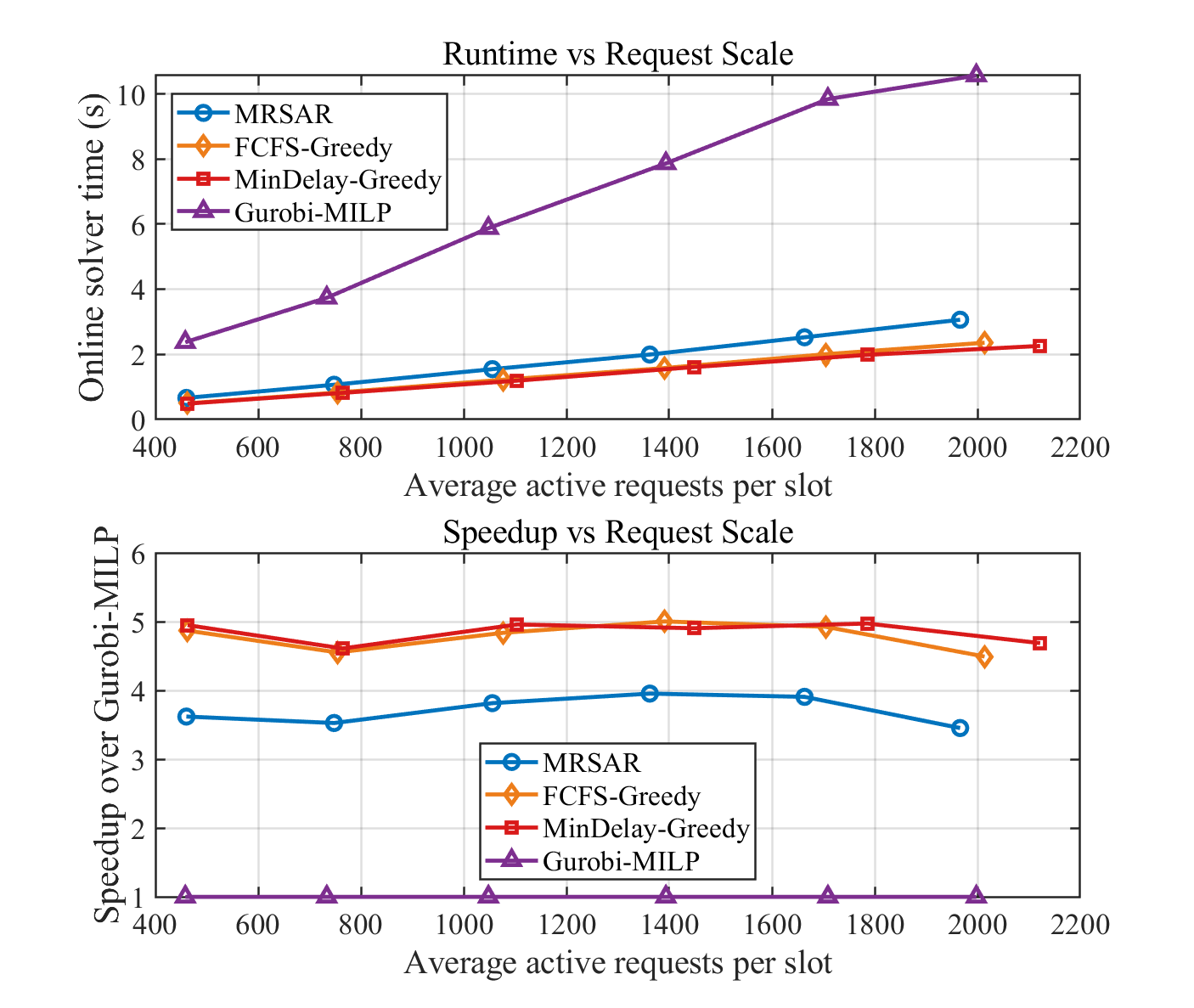}
\caption{Runtime scalability and speedup over Gurobi-MILP under different request scales.}
\label{fig:runtime_speedup}
\end{figure}

Fig.~\ref{fig:runtime_speedup} evaluates runtime scalability under
different request scales. As shown in the upper panel, the runtime of
Gurobi-MILP increases from approximately 2.6~s to 10.6~s as the average
number of active requests per slot grows. In comparison, the runtime of
MRSAR increases more moderately from about 0.7~s to 3.2~s. The two greedy
baselines incur slightly lower runtime than MRSAR, while remaining in the
same computational order. This indicates that the additional risk- and
scarcity-aware refinement introduces limited overhead relative to simple
greedy construction. Combined with the preceding success and failure
results, MRSAR provides a more favorable tradeoff between computational
cost and orchestration quality.

The lower panel shows that MRSAR achieves an approximately
$3.5$--$4.0\times$ speedup over Gurobi-MILP across the evaluated request
scales. The speedup varies moderately because MILP runtime depends not only
on the request scale, but also on the feasible-action structure, resource
contention, and the resulting branch-and-bound process. Although the greedy
baselines achieve somewhat higher speedups, their lower computational cost
is accompanied by larger QoS-success gaps. All runtime results are measured
using an unoptimized single-machine research prototype and are therefore
intended as a relative computational comparison rather than the latency of
an engineering-grade controller. Overall, MRSAR scales more effectively than
directly solving the full MILP while retaining substantially better
orchestration quality than the greedy baselines.

\subsection{Ablation Study and TEMP Sensitivity}
\label{subsec:ablation_temp}

\begin{figure}[!t]
\centering
\includegraphics[
width=\simfigwidth,
height=\simfigheight,
keepaspectratio
]{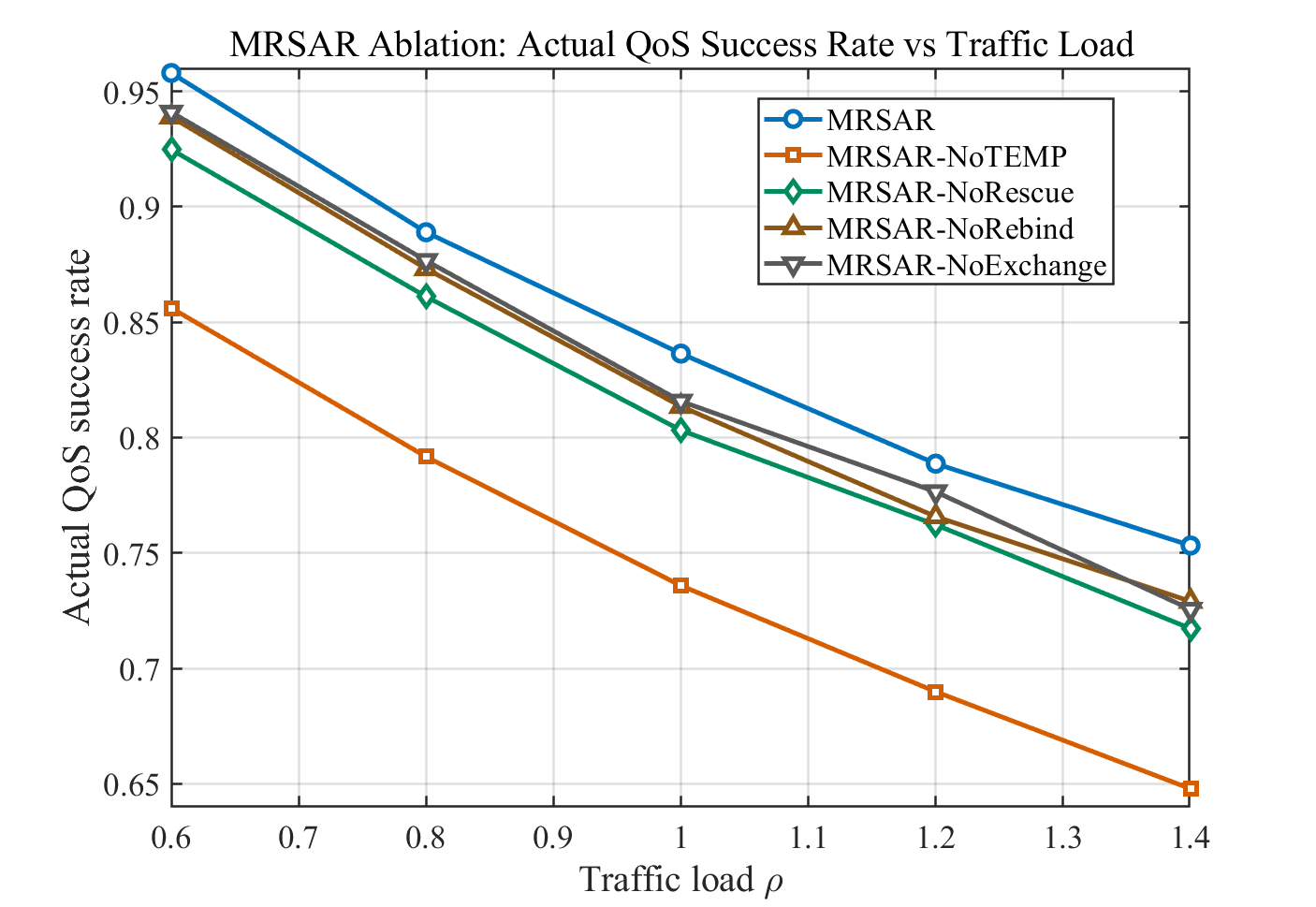}
\caption{Ablation study of MRSAR.}
\label{fig:ablation}
\end{figure}

Fig.~\ref{fig:ablation} evaluates the contribution of the main components in MRSAR. Removing TEMP causes the most significant degradation, which is consistent with the TEMP utilization behavior in Fig.~\ref{fig:temp_buffer_utilization}. Without TEMP, requests that cannot be served immediately lose the opportunity to exploit future data-link availability, leading to a clear success-ratio loss. Removing Rescue, Rebind, or Exchange operations also degrades performance, but the degradation is generally smaller than that caused by disabling TEMP. This indicates that TEMP provides the main temporal elasticity, while action refinement improves how these delayed opportunities are converted into successful deliveries.

\begin{figure}[!t]
\centering
\includegraphics[
width=\simfigwidth,
height=\simfigheight,
keepaspectratio
]{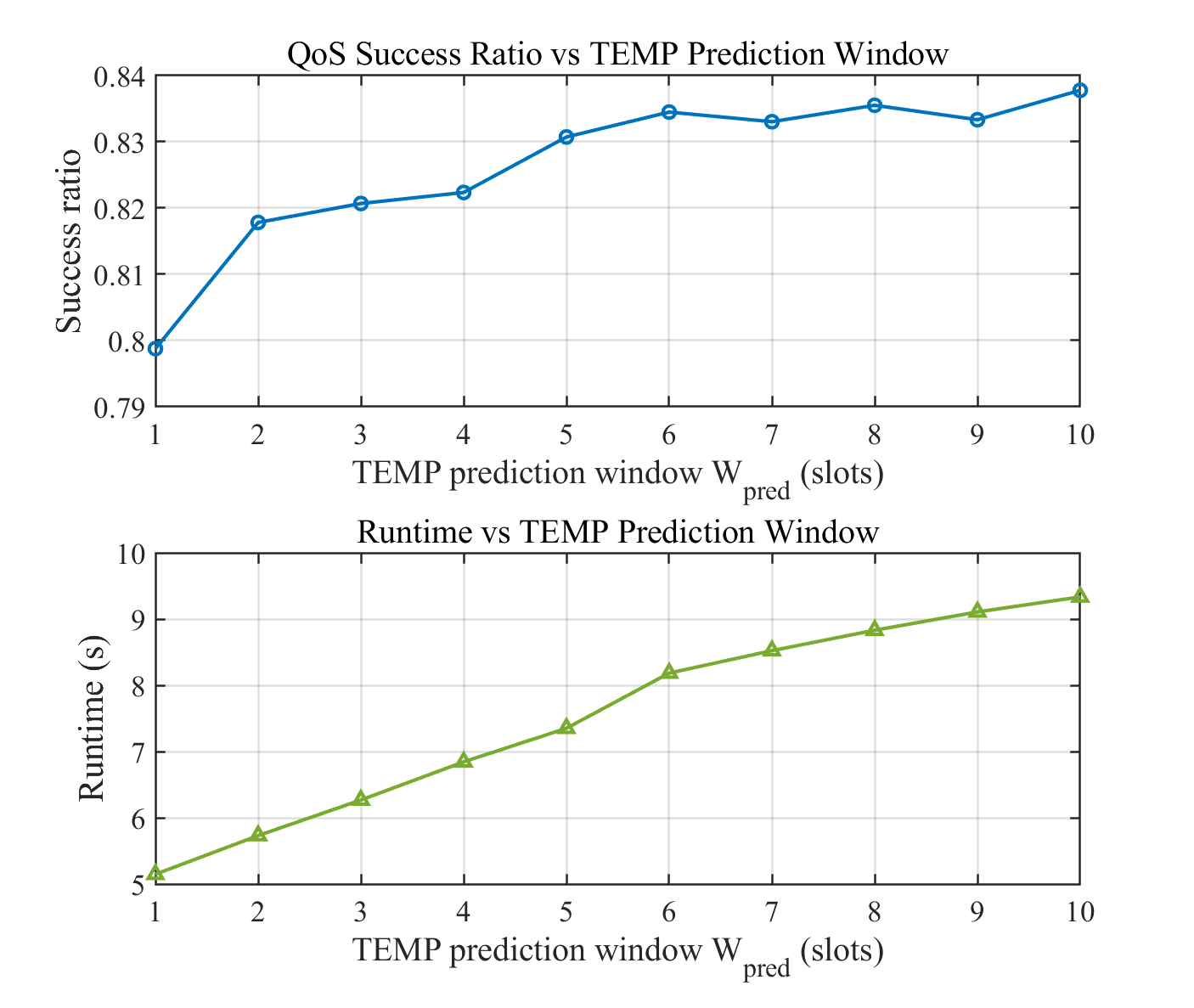}
\caption{Sensitivity of MRSAR to the TEMP look-ahead window length.}
\label{fig:temp_window_sensitivity}
\end{figure}

Fig.~\ref{fig:temp_window_sensitivity} examines the sensitivity of MRSAR to the TEMP look-ahead window length. A short window limits the visibility of future data-link opportunities and reduces the value of TEMP admission. As the window length increases, MRSAR can identify more future realization opportunities and improve the success-oriented objective. The gain gradually saturates when the window becomes sufficiently long, because additional future slots provide less marginal information while increasing computational overhead. This trend supports the use of a finite rolling prediction window: limited future information is useful for online orchestration, but an excessively long window is unnecessary.

Overall, the simulation results show that MRSAR achieves a favorable balance among QoS success, delay control, failure reduction, near-optimality, and runtime scalability. Its main performance gain comes from treating TEMP as a schedulable temporal resource. The model-induced risk value and scarcity-aware resource signals further determine how these temporal opportunities should be selected, preserved, and realized during rolling online orchestration.

\section{Conclusion}
\label{sec:conclusion}

This paper investigated SFC-aware online access-side aggregated
data-link orchestration for SDN/NFV-enabled civil aviation
SAGINs. To support heterogeneous cockpit and cabin services
under highly dynamic communication conditions, heterogeneous
A2A, A2G, and A2S data links were abstracted as aggregated
bearer resources, and request-level access orchestration was
formulated over joint access-anchor and bearer-mode selection.
In addition to immediate service through current bearer resources,
TEMP was introduced as a temporal elastic mapping and parking
mechanism to preserve selected requests for future realization
within a finite look-ahead window. Based on this spatial-temporal
decision structure, four request-level actions, namely NOW,
TEMP, REJECT, and DROP, were used to describe the rolling
online orchestration process.

A lexicographic rolling-slot MILP was developed to capture the
main operational priorities of the considered access orchestration
problem. The model first maximizes service success under QoS
constraints, then minimizes normalized access-orchestration delay,
and finally controls residual TEMP-related risk, including DROP,
residence urgency, and TEMP buffer pressure. Since repeatedly
solving the full binary model is costly for fine-grained online
control, MRSAR was proposed as a model-induced risk and
scarcity-aware refinement algorithm. MRSAR derives deferred-
realization risk from the success-prioritized objective and resource
scarcity signals from bearer-capacity and TEMP-buffer constraints.
These signals are used for valuation-based feasible solution
construction and bounded, feasibility-preserving neighborhood
refinement, allowing the algorithm to preserve the main decision
logic of the MILP while reducing online computational cost.

Simulation results showed that MRSAR remains close to the
Gurobi-MILP reference in service success and consistently
outperforms arrival-order and delay-myopic greedy baselines.
The results also showed that MRSAR maintains normalized delay
close to the MILP reference, controls REJECT and DROP failures,
and uses TEMP as a managed temporal resource rather than a
passive buffer. Runtime and optimality-gap evaluations further
confirmed that MRSAR provides a favorable quality-complexity
tradeoff for rolling online orchestration. The ablation and window-
sensitivity results indicated that both TEMP-aware preservation
and model-induced risk-scarcity refinement contribute to the
observed performance gains. Future work will integrate practical
traffic and link-state prediction modules into the proposed
framework and evaluate the robustness of TEMP-aware
orchestration under prediction uncertainty.

\bibliographystyle{IEEEtran}
\bibliography{references}

@article{liu2018sagin,
author  = {J. Liu and Y. Shi and Z. M. Fadlullah and N. Kato},
title   = {Space-Air-Ground Integrated Network: A Survey},
journal = {IEEE Communications Surveys \& Tutorials},
volume  = {20},
number  = {4},
pages   = {2714--2741},
year    = {2018},
doi     = {10.1109/COMST.2018.2841996}
}

@article{cheng2022service_sagin,
author  = {N. Cheng and J. He and Z. Yin and C. Zhou and H. Wu and F. Lyu and H. Zhou and X. Shen},
title   = {6G Service-Oriented Space-Air-Ground Integrated Network: A Survey},
journal = {Chinese Journal of Aeronautics},
volume  = {35},
number  = {9},
pages   = {1--18},
year    = {2022},
doi     = {10.1016/j.cja.2021.12.013}
}

@article{zhou2023aerospace,
author  = {D. Zhou and M. Sheng and J. Li and Z. Han},
title   = {Aerospace Integrated Networks Innovation for Empowering 6G: A Survey and Future Challenges},
journal = {IEEE Communications Surveys \& Tutorials},
volume  = {25},
number  = {2},
pages   = {975--1019},
year    = {2023},
doi     = {10.1109/COMST.2023.3245614}
}

@article{azari2022ntn,
author  = {M. M. Azari and S. Solanki and S. Chatzinotas and O. Kodheli and H. Sallouha and A. Colpaert and J. F. Mendoza Montoya and S. Pollin and A. Haqiqatnejad and A. Mostaani and E. Lagunas and B. Ottersten},
title   = {Evolution of Non-Terrestrial Networks From 5G to 6G: A Survey},
journal = {IEEE Communications Surveys \& Tutorials},
volume  = {24},
number  = {4},
pages   = {2633--2672},
year    = {2022},
doi     = {10.1109/COMST.2022.3199901}
}

@article{bilen2022aeronautical,
author  = {T. Bilen and H. Ahmadi and B. Canberk and T. Q. Duong},
title   = {Aeronautical Networks for In-Flight Connectivity: A Tutorial of the State-of-the-Art and Survey of Research Challenges},
journal = {IEEE Access},
volume  = {10},
pages   = {20053--20079},
year    = {2022},
doi     = {10.1109/ACCESS.2022.3151658}
}

@article{baltaci2021aerial_networks,
author  = {A. Baltaci and E. Dinc and M. Ozger and A. Alabbasi and C. Cavdar and D. Schupke},
title   = {A Survey of Wireless Networks for Future Aerial Communications},
journal = {IEEE Communications Surveys \& Tutorials},
volume  = {23},
number  = {4},
pages   = {2833--2884},
year    = {2021},
doi     = {10.1109/COMST.2021.3103044}
}

@article{liang2024resource_sagin,
author  = {H. Liang and Z. Yang and G. Zhang and H. Hou},
title   = {Resource Allocation for Space-Air-Ground Integrated Networks: A Comprehensive Review},
journal = {Journal of Communications and Information Networks},
volume  = {9},
number  = {1},
pages   = {1--23},
year    = {2024},
doi     = {10.23919/JCIN.2024.10494938}
}

@article{zhang2022multi_domain_sagin,
author  = {P. Zhang and C. Wang and N. Kumar and L. Liu},
title   = {Space-Air-Ground Integrated Multi-Domain Network Resource Orchestration Based on Virtual Network Architecture: A DRL Method},
journal = {IEEE Transactions on Intelligent Transportation Systems},
volume  = {23},
number  = {3},
pages   = {2798--2808},
year    = {2022}
}

@article{zhang2024ai_sagin,
author  = {P. Zhang and N. Chen and S. Shen and S. Yu and N. Kumar and C.-H. Hsu},
title   = {AI-Enabled Space-Air-Ground Integrated Networks: Management and Optimization},
journal = {IEEE Network},
volume  = {38},
number  = {2},
pages   = {186--192},
year    = {2024},
doi     = {10.1109/MNET.131.2200477}
}

@article{kreutz2015sdn,
author  = {D. Kreutz and F. M. V. Ramos and P. E. Verissimo and C. E. Rothenberg and S. Azodolmolky and S. Uhlig},
title   = {Software-Defined Networking: A Comprehensive Survey},
journal = {Proceedings of the IEEE},
volume  = {103},
number  = {1},
pages   = {14--76},
year    = {2015},
doi     = {10.1109/JPROC.2014.2371999}
}

@article{mijumbi2016nfv,
author  = {R. Mijumbi and J. Serrat and J.-L. Gorricho and N. Bouten and F. De Turck and R. Boutaba},
title   = {Network Function Virtualization: State-of-the-Art and Research Challenges},
journal = {IEEE Communications Surveys \& Tutorials},
volume  = {18},
number  = {1},
pages   = {236--262},
year    = {2016},
doi     = {10.1109/COMST.2015.2477041}
}

@article{afolabi2018network_slicing,
author  = {I. Afolabi and T. Taleb and K. Samdanis and A. Ksentini and H. Flinck},
title   = {Network Slicing and Softwarization: A Survey on Principles, Enabling Technologies, and Solutions},
journal = {IEEE Communications Surveys \& Tutorials},
volume  = {20},
number  = {3},
pages   = {2429--2453},
year    = {2018},
doi     = {10.1109/COMST.2018.2815638}
}

@misc{halpern2015sfc_arch,
author       = {J. Halpern and C. Pignataro},
title        = {Service Function Chaining (SFC) Architecture},
howpublished = {IETF RFC 7665},
year         = {2015},
doi          = {10.17487/RFC7665}
}

@article{bhamare2016sfc_survey,
author  = {D. Bhamare and R. Jain and M. Samaka and A. Erbad},
title   = {A Survey on Service Function Chaining},
journal = {Journal of Network and Computer Applications},
volume  = {75},
pages   = {138--155},
year    = {2016},
doi     = {10.1016/j.jnca.2016.09.001}
}

@article{hantouti2020sfc_5gb,
author  = {H. Hantouti and N. Benamar and T. Taleb},
title   = {Service Function Chaining in 5G \& Beyond Networks: Challenges and Open Research Issues},
journal = {IEEE Network},
volume  = {34},
number  = {4},
pages   = {320--327},
year    = {2020},
doi     = {10.1109/MNET.001.1900554}
}

@article{li2022cost_sfc_sagin,
author  = {J. Li and W. Shi and H. Wu and S. Zhang and X. Shen},
title   = {Cost-Aware Dynamic SFC Mapping and Scheduling in SDN/NFV-Enabled Space-Air-Ground-Integrated Networks for Internet of Vehicles},
journal = {IEEE Internet of Things Journal},
volume  = {9},
number  = {8},
pages   = {5824--5838},
year    = {2022},
doi     = {10.1109/JIOT.2021.3058250}
}

@article{zhang2022sfc_sagin,
author  = {P. Zhang and P. Yang and N. Kumar and M. Guizani},
title   = {Space-Air-Ground Integrated Network Resource Allocation Based on Service Function Chain},
journal = {IEEE Transactions on Vehicular Technology},
volume  = {71},
number  = {7},
pages   = {7730--7738},
year    = {2022},
doi     = {10.1109/TVT.2022.3165145}
}

@article{zhang2023fl_sfc_sagin,
author  = {P. Zhang and Y. Zhang and N. Kumar and M. Guizani},
title   = {Dynamic SFC Embedding Algorithm Assisted by Federated Learning in Space-Air-Ground-Integrated Network Resource Allocation Scenario},
journal = {IEEE Internet of Things Journal},
volume  = {10},
number  = {11},
pages   = {9308--9318},
year    = {2023},
doi     = {10.1109/JIOT.2022.3222200}
}

@article{jia2025sfc_scheduling,
author  = {Z. Jia and Y. Cao and L. He and Q. Wu and Q. Zhu and D. Niyato and Z. Han},
title   = {Service Function Chain Dynamic Scheduling in Space-Air-Ground Integrated Networks},
journal = {IEEE Transactions on Vehicular Technology},
volume  = {74},
number  = {7},
pages   = {11235--11249},
year    = {2025},
doi     = {10.1109/TVT.2025.3543259}
}

@article{du2021dynamic_graph,
author  = {B. Du and X. Di and D. Liu and H. Zhang},
title   = {Dynamic Graph Optimization and Performance Evaluation for Delay-Tolerant Aeronautical Ad Hoc Network},
journal = {IEEE Transactions on Communications},
volume  = {69},
number  = {9},
pages   = {6018--6036},
year    = {2021},
doi     = {10.1109/TCOMM.2021.3085898}
}

@article{cui2021minimum_delay_aanet,
author  = {J. Cui and D. Liu and J. Zhang and H. Yetgin and S. X. Ng and R. G. Maunder and L. Hanzo},
title   = {Minimum-Delay Routing for Integrated Aeronautical Ad Hoc Networks Relying on Real Flight Data in the North-Atlantic Region},
journal = {IEEE Open Journal of Vehicular Technology},
volume  = {2},
pages   = {310--320},
year    = {2021},
doi     = {10.1109/OJVT.2021.3089543}
}

@article{zhang2022multiobjective_aanet,
author  = {J. Zhang and D. Liu and S. Chen and S. X. Ng and R. G. Maunder and L. Hanzo},
title   = {Multiple-Objective Packet Routing Optimization for Aeronautical Ad-Hoc Networks},
journal = {IEEE Transactions on Vehicular Technology},
volume  = {72},
number  = {1},
pages   = {1002--1016},
year    = {2023},
doi     = {10.1109/TVT.2022.3202689}
}

@book{kleinrock1975queueing,
author    = {L. Kleinrock},
title     = {Queueing Systems, Volume 1: Theory},
publisher = {Wiley},
address   = {New York, NY, USA},
year      = {1975}
}

@article{ojijo2020slice_admission,
author  = {M. O. Ojijo and O. E. Falowo},
title   = {A Survey on Slice Admission Control Strategies and Optimization Schemes in 5G Network},
journal = {IEEE Access},
volume  = {8},
pages   = {14977--14990},
year    = {2020},
doi     = {10.1109/ACCESS.2020.2967626}
}

@article{gallager1977minimum_delay,
author  = {R. G. Gallager},
title   = {A Minimum Delay Routing Algorithm Using Distributed Computation},
journal = {IEEE Transactions on Communications},
volume  = {25},
number  = {1},
pages   = {73--85},
year    = {1977},
doi     = {10.1109/TCOM.1977.1093731}
}

@misc{gurobi2026,
author       = {{Gurobi Optimization, LLC}},
title        = {Gurobi Optimizer Reference Manual},
year         = {2026},
note         = {Version 13.0}
}

@article{li2018drl_slicing,
author  = {R. Li and Z. Zhao and Q. Sun and C. I and C. Yang and X. Chen and M. Zhao and H. Zhang},
title   = {Deep Reinforcement Learning for Resource Management in Network Slicing},
journal = {IEEE Access},
volume  = {6},
pages   = {74429--74441},
year    = {2018},
doi     = {10.1109/ACCESS.2018.2881964}
}

@article{liu2021drl_aanet,
author  = {D. Liu and J. Cui and J. Zhang and C. Yang and L. Hanzo},
title   = {Deep Reinforcement Learning Aided Packet-Routing for Aeronautical Ad-Hoc Networks Formed by Passenger Planes},
journal = {IEEE Transactions on Vehicular Technology},
volume  = {70},
number  = {5},
pages   = {5166--5171},
year    = {2021},
doi     = {10.1109/TVT.2021.3074015}
}

\end{document}